%% file: main.tex
\documentclass[sigconf]{acmart}

\acmConference{}{}{}
\acmDOI{}
\acmISBN{}

\AtBeginDocument{%
  \providecommand\BibTeX{{%
    \normalfont B\kern-0.5em{\scshape i\kern-0.25em b}\kern-0.8em\TeX}}}

\setcopyright{rightsretained}

\usepackage{svg}
\usepackage{multirow}
\usepackage{graphicx}
\usepackage{caption}
\usepackage{subcaption}
\usepackage{algorithmic}
\DeclareMathOperator*{\argmin}{arg\,min}

\DeclareMathOperator{\E}{\mathbb{E}}
\newcommand{\paragraphX}[1]{\vskip 4pt \noindent \textbf{\textit{#1}} \hskip .05in}
\usepackage{todonotes}

\newenvironment{smitemize}{
\begin{itemize}
  \setlength{\topsep}{-3pt}
  \setlength{\itemsep}{1pt}
  \setlength{\parskip}{0pt}
  \setlength{\parsep}{0pt}
  \setlength{\leftmargin}{1em}  \setlength{\itemindent}{-4pt}
}{\end{itemize}}

\begin{document}

\title{Detecting Adversarial Patches with Class Conditional Reconstruction Networks}


\author{Perry Deng}
\affiliation{\institution{Rochester Institute of Technology}}
\email{perry.deng@mail.rit.edu}

\author{Mohammad Saidur Rahman}
\affiliation{\institution{Rochester Institute of Technology}}
\email{saidur.rahman@mail.rit.edu}

\author{Matthew Wright}
\affiliation{\institution{Rochester Institute of Technology}}
\email{matthew.wright@rit.edu}

\renewcommand{\shortauthors}{Anonymous}

\input{0_abstract}


\maketitle
\newcommand\nnfootnote[1]{
    \begin{NoHyper}
    \renewcommand\thefootnote{}\footnote{#1}
    \addtocounter{footnote}{-1}
\end{NoHyper}
}
\nnfootnote{This material is based upon work supported by the National Science Foundation under Award No. 1816851.}
\subsection*{Resources} Source code: \url{https://github.com/PerryXDeng/detecting-adversarial-patches-capsule-networks}

\input{1_intro}

\input{2_related_work}

\input{3_threat_model}

\input{4_architecture}
\input{5_exp_setup}
\input{6_exp_results}
\input{7_conclusion}


\bibliographystyle{ACM-Reference-Format}
\bibliography{refs}

\balance
\newpage
\appendix
\input{99_patches}

\end{document}

%% file: 0_abstract.tex
\begin{abstract}
Defending against physical adversarial attacks is a rapidly growing topic in deep learning and computer vision. Prominent forms of physical adversarial attacks, such as overlaid adversarial patches and objects, share similarities with digital attacks, but are easy for humans to notice. This leads us to explore the hypothesis that adversarial detection methods, which have been shown to be ineffective against adaptive digital adversarial examples, can be effective against these physical attacks. We use one such detection method based on autoencoder architectures, and perform adversarial patching experiments on MNIST, SVHN, and CIFAR10 against a CNN architecture and two CapsNet architectures. We also propose two modifications to the EM-Routed CapsNet architecture, {\em Affine Voting} and {\em Matrix Capsule Dropout}, to improve its classification performance. Our investigation shows that the detector retains some of its effectiveness even against adaptive adversarial patch attacks. In addition, detection performance tends to decrease among all the architectures with the increase of dataset complexity.

\end{abstract}

%% file: 1_intro.tex
\section{Introduction}
\label{intro}




Deep-learning (DL) systems have shown tremendous success in complex tasks such as image recognition~\cite{krizhevsky2012imagenet}, speech recognition~\cite{hinton2012deep}, object detection, classification, tracking~\cite{geiger2012we, lillicrap2015continuous}, and so on. Much research, however, has shown that DL systems are vulnerable to {\em digital adversarial examples}, carefully crafted inputs to cause unintended behaviors. One method of defending against these attacks is to detect the attacks and enable the system to formulate a new action based on the detection. Even though a number of methods have been proposed for detecting adversarial examples~\cite{metzen2017detecting, grosse2017statistical} in the digital domain, these methods have shown to be ineffective against an adaptive attacker~\cite{carlini2017bypass} who targets a DL system to cause misclassification and evade detection. 

Growing research shows that similar vulnerabilities exist in the physical domain as well~\cite{lee2019physical,chou2018sentinet}.
Physical adversarial attacks are created to be robust to variance in scene and
sensor factors such as lighting, rotation, and distance, rather than merely making them effective in a single example image. For example, attackers can craft {\em adversarial patches} and place them anywhere within the field of view of an object detector to suppress object detection~\cite{lee2019physical}. This is achieved through Expectation over Transformation (EOT)~\cite{athalye2017synthesizing}. EOT simulates a distribution of locations, orientations, and photometric properties of the patch, and then optimizes the attack across the dataset with the distribution applied. Unlike digital adversarial examples, adversarial patches have to occupy a continuous area in input images. This property makes them easy to be spot for humans, for example in the cases of toaster patches~\cite{brown2017adversarial} or printed t-shirts~\cite{xu2019adversarial}. Adversarial patches are cheap to craft, and unlike digital adversarial examples, can work in many scenarios. Adversarial patch attacks on time-sensitive safety-critical systems, such as a self-driving car, can jeopardize users as well as the general public. Detecting adversarial patch attacks, however, has received considerably less attention until recently.

We note that the only differences between the generation of adversarial patches and digital adversarial examples are that: i) adversarial patches are optimized over a distribution of input images rather than a single image, ii) adversarial patches are area constrained rather than norm constrained.  We thus hypothesize that \emph{neural networks under attack by adversarial patches are behaviorally similar to neural networks under attack by digital adversarial examples, and methods to detect adversarial perturbations can be transferred over to detection of adversarial patches}.

Intuitively, it should also be easier to automate the detection of adversarial patches compared to digital adversarial examples, as humans can easily spot them. To the best of our knowledge, SentiNet~\cite{chou2018sentinet} is the only prior work done on detecting adversarial patches. SentiNet detects and segments adversarial objects from input by leveraging the high salience and scene-transferability of adversarial patches. 

In this paper, we adapt a mechanism originally proposed to detect digital adversarial perturbations~\cite{qin2019detecting} and empirically examine its effectiveness against simulated physical adversarial attacks.
%
In particular, we examine the effectiveness of class-conditional reconstructions~\cite{qin2019detecting} -- which is lackluster in the digital domain -- against adversarial patches. We show that unlike digital adversarial examples, adaptive attacks using adversarial patches do not neutralize the detector. We have implemented the detection approach on a modified AlexNet as well as on a \emph{Capsule Network (CapsNet)}, a recently proposed and arguably powerful neural architecture~\cite{sabour2017dynamic,hinton2018matrix}. CapsNets use layers of vector- or matrix-valued capsules that represent the poses and probabilities of objects and can also represent part-whole relationships. Similar to how layers of neurons can represent the extraction of higher-level features from lower-level features in a standard neural network, layers of capsules can represent the extraction of higher-level objects from lower-level objects, shapes, or lines. In addition, we propose an improved CapsNet architecture incorporating {\em Dropout} and {\em Affine Voting} to enhance the test set classification performance.

In summary, our contributions are as follows: 
\begin{smitemize}
    \item We are the first to explore the effectiveness of a digital adversarial example detection method in the adversarial patch domain, and demonstrate it on three standard datasets: MNIST, SVHN, and CIFAR10.
    \item We are the first to show the effectiveness of CapNets to classify and detect adversarially patched input.
    \item We propose an improved Matrix Capsule Network with {\em capsule-dropout} and {affine voting} and empirically show their improved test accuracy on the smallNORB dataset.
\end{smitemize}

%% file: 2_related_work.tex
\section{Related Work}
\label{rel_work}


\paragraphX{Digital Adversarial Examples.}
Digital adversarial examples are inputs slightly perturbed by attackers to cause unintended behaviors in deep-learning (DL) system. The creation of adversarial examples in the digital domain has been extensively studied using convolutional neural networks (CNNs)~\cite{szegedy2013intriguing, moosavi2016deepfool, carlini2017towards}. These digital domain attacks typically assume that the threat actors can modify pixels of the input images at will, and seek to optimize an adversarial objective that would lead to misclassification subject to some constraints on the amount of modification.

While many methods have been proposed to detect digitally perturbed adversarial input~\cite{feinman2017detecting, lu2017quantized, pang2017robust, xu2018feature, meng2017magnet}, none of these approaches have been shown to be robust against an adaptive digital domain attacker aware of the defense mechanism~\cite{carlini2017bypass, carlini2017magnet, yu2019weakness, qin2019detecting}. Some of the more promising approaches, however, force the attacker to employ digital perturbations that are more noticeable to human observers~\cite{feinman2017detecting, xu2018feature, yu2019weakness, qin2019detecting}.

%
%
\paragraphX{Physical Adversarial Attacks.}
\label{related_adv}
Unlike digital adversarial examples, physical adversarial attacks do not assume that an attacker can manipulate any pixel on a particular input image. In the computer vision setting, physical adversarial attacks produce an object (which we generically refer to as a \emph{patch}, as objects are but patches of pixels in a 2d image) that tricks the image classifier into seeing a particular type of object. These patches are designed to be robust to variance in scene and sensor factors such as lighting, rotation, and distance, rather than merely effective in a single example image. This is achieved through Expectation over Transformation (EOT), a method first proposed by Athalye et al.~\cite{athalye2017synthesizing}.

EOT simulates a distribution of locations, orientations, and photometric properties of the patch and then optimizes the attack across all the data with the simulated distribution applied. Patches can be designed to modify a specific target object to be classified or to appear with any other objects. Attack that modifies the target objects can directly manipulate the surfaces of the objects using technologies such as 3D printing. One example of this attack produces an object that looks like a turtle to human but is classified as a rifle by the neural network~\cite{athalye2017synthesizing}. The second type of physical attacks pose more threat to an undefended system, as they do not assume the ability to modify the target object. Rather they place adversarial objects or patches in the scene intended to interfere with the neural networks. Examples of these attacks include i) hiding people from detection through adversarial patches on t-shirts~\cite{thys2019fooling}, ii) causing neural algorithms to lose track of vehicles through adversarial patches~\cite{jia2019fooling}.

The investigation of this paper is {\em adversarial patch} centric as it requires relatively low cost to perform such attack. Adversarial patches can fool a trained DL system to misclassify~\cite{brown2017adversarial} when applied to a sticker or other objects. Unlike constraining the magnitude of pixel change of digital adversarial examples, this class of attacks constrains the payloads to be patches that cover a continuous area, which makes it easy for humans to recognize.

\paragraphX{Capsule Networks.}
\label{related_cap}
Capsule networks are a recently proposed alternative class of neural network architecture~\cite{hinton2011transforming, sabour2017dynamic, hinton2018matrix, kosiorek2019stacked} that use layers of vector- or matrix-valued capsules, which represent the poses and probability of objects. Similar to how layers of neurons can represent the extraction of higher-level features from lower-level features in a standard neural network, layers of capsules can represent the extraction of higher-level objects from lower-level objects, shapes, or lines. 
A typical DNN consists of layers of scalar-valued neurons that are connected together using linear combinations and non-linear activation functions.

By contrast,
in a capsule network, layers of capsules vote on the poses and probability of higher level capsules, and the probabilities of their existence are determined by a procedure known as {\em Routing by Agreement}~\cite{sabour2017dynamic,hinton2018matrix}.
Routing by Agreement eliminates the need for pooling, a dimensionality reduction technique commonly used in a CNN. Two approaches to this routing have been proposed, both of which have the theoretical property of geometric equivariance. This makes them robust to geometric transformations of objects without the need to memorizing different viewpoints and placement of objects, and reduces the need for training data~\cite{hinton2018matrix}. The first approach, {\em Dynamic Routing}~\cite{sabour2017dynamic}, measures agreement between capsules using the dot product between votes and capsules that are voted on, and use an $ad-hoc$ squashing function to normalize the final vector capsule output with the magnitude being its probability. The Dynamic Routing implementation has shown state-of-the-art performance in tasks such as medical image segmentation~\cite{lalonde2018capsules}. The second approach, {\em EM Routing}~\cite{hinton2018matrix}, measures agreement between votes using Gaussian clusters estimated with a modified {\em Expectation-Maximization} (EM) algorithm.

The EM Routing implementation allows the use of matrix capsules, and is shown to generalize better to novel viewpoints while requiring less data to train~\cite{hinton2018matrix} on the smallNORB dataset, compared to CNN. Our architecture is based on matrix capsules with EM Routing, the second implementation. Our proposed improvements to the architecture are discussed in Section~\ref{modifications}.

\paragraphX{Digital Adversarial Example Detection with Reconstruction Networks.}
\label{decoder}
MagNet~\cite{meng2017magnet} shows that autoencoders have the potential to counter adversarially perturbed input, which are outside of the manifold boundary of the normal input, by reconstructing the input as if they are inside the normal manifold. This allows the detection of digital adversarial examples when the class-conditioned reconstruction is far from the provided input, a method first proposed by Qin et al.~\cite{qin2019detecting}. The method feeds the class-conditioned latent variables of the image classifier into a decoder network turning the architecture into an autoencoder. The decoder output then outputs a tensor that attempts to reconstruct the input images given latent variables that represent {\em features} of the predicted class instance. In a capsule network, the latent variables are the class capsule vectors or matrices in the final layer of the classifier. The class conditional decoder can also be attached to a convolutional network, albeit with some modifications to the penultimate layer in the CNN~\cite{qin2019detecting}. The $L_2$ distance between the reconstructed input $x_{recon}$ and the actual input $x$, is minimized when training the network. The adversarial detection function $f_{detect}$, which outputs $1$ when the input is considered to be adversarial, and $0$ otherwise, is defined as
\begin{gather}
    f_{detect}(x) = H(\Delta_{recon} - \alpha_{crit}),
    \label{eq:detection}
\end{gather}
where $H$ is the Heaviside step function, $\Delta_{recon} = \frac{\|x_{recon} - x\|_{2}}{height_x*width_x}$ is the reconstruction loss, and $\alpha_{crit}$ is the critical threshold. In our experiments, we pick $\alpha_{crit}$ to be at the 95th percentile of reconstruction losses on test set images, so that the false positive rate is controlled at $5\%$.
%

%% file: 3_threat_model.tex
\section{Threat Model}
\label{threat}

\begin{figure}[t!]
  \centering
  \includegraphics[scale=0.375]{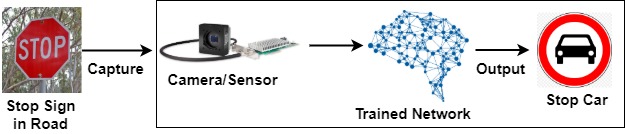}
  \caption{A deep-learning-based computer-vision system (CVS), namely an autonomous vehicle, which should: i) capture an image of the stop sign from the physical world, ii) feed the image into the trained deep neural network, and iii) provide the classification output to a decision engine that stops the car.}
 \vspace{-0.5cm}
  \label{fig:system}
\end{figure}

\subsection{Physical Adversarial Attacks}
From an attacker's point of view, the example of self-driving car in Figure~\ref{fig:system} can be exploited to behave in an unintended and potentially dangerous manner by taking over the steering or power system (as was done in a non-autonomous vehicle~\cite{greenberg2018jeep}), poisoning the neural network training data to plant backdoors~\cite{liu2017trojaning}, manipulating the capturing or preprocessing procedure to modify images to have human-imperceptible perturbations that target neural networks~\cite{szegedy2013intriguing, moosavi2016deepfool, carlini2017towards}, or crafting and placing physical adversarial objects in the environment that target the neural networks~\cite{kurakin2016adversarial, athalye2017synthesizing, brown2017adversarial}. While attacks have been examined for all of those vulnerabilities, placement of crafted adversarial payloads in the physical environment assumes the least amount of privileged access to the targeted systems. Sophisticated adversarial attacks have been proposed to create physical adversarial patches and objects that are robust to environmental variances, and can exploit different parts of the computer vision pipeline such as object classification~\cite{athalye2017synthesizing, brown2017adversarial}, object detection~\cite{chen2018robust, lee2019physical}, and object tracking~\cite{jia2019fooling}.

\subsection{Attack Detection}
While physical payloads created by the aforementioned techniques are in general more noticeable to humans than digital adversarial perturbations, a system armed with an automatic detector of these attacks can mitigate risks, as they do not require trained eyes to identify such a situation. For example, a self-driving car can relegate its steering to a licensed driver once it detects adversarial objects in the environment, thereby reducing the chances of unexpected maneuvers and traffic collisions, without requiring professionals trained at spotting such attacks. 

In this paper, we concentrate on detecting the presence of adversarial patches or objects in the object classification pipeline, as it is relatively easy to demonstrate and measure the results. It is also possible to adapt our model to a more complex computer vision system (CVS). To evaluate the effectiveness of our detection method, we assume the worst-case scenario for adversarial patch or object attacks, in which the threat actors possess access to parameters of the neural network model in the CVS, even though they do not possess privileged access to the instance of the targeted system. This white-box scenario is a reasonable assumption, as it has been shown to be possible for an attacker to transfer payloads trained against a substitute model to the targeted black-box model that attackers do not have access to~\cite{papernot2016transferability}. With this assumption, we then evaluate the effectiveness of the detection method by analyzing its robustness to two attacks:

\begin{enumerate}
    \item A naive white-box attack that focuses on creating a strong attack against the classifier with no regard for the detection in place.
    \item An adaptive white-box attack that attempts to fool both the classifier \textit{and} the detector.
\end{enumerate}

For detection to be considered effective, it needs to not only effectively detect strong naive attacks, but also maintain a reasonable performance against adaptive attacks. This evaluation methodology is identical to the ones used by Biggio et al.~\cite{biggio2013evasion} and Carlini and Wagner~\cite{carlini2017bypass}. The two attacks used for evaluation are described in detail in Section~\ref{adv_patching}.

%% file: 4_architecture.tex

\begin{figure}[!t]
  \centering
  \includegraphics[scale=0.25]{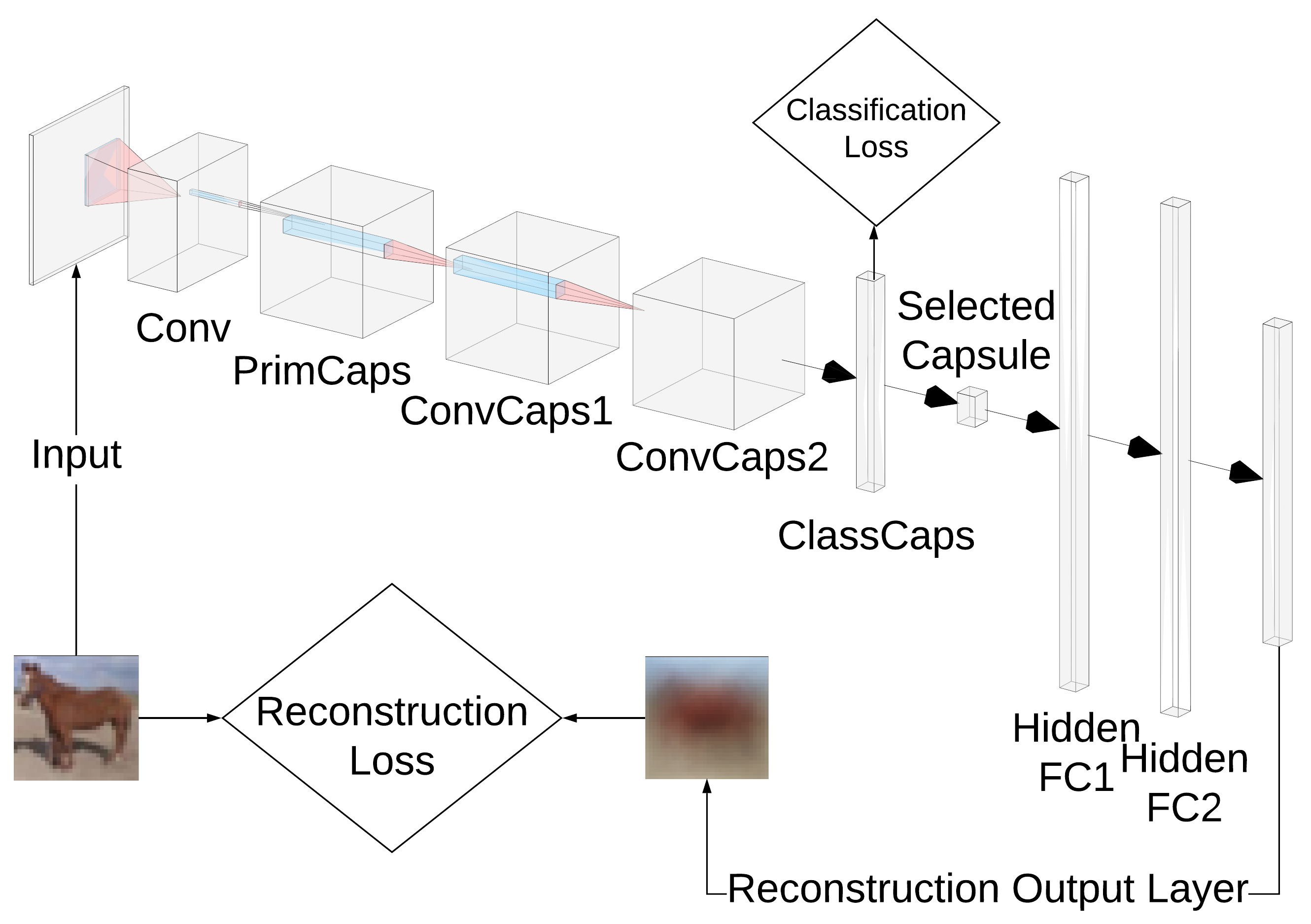}
  \caption{Overview of the CapsNet Detection Architecture.}
  \label{fig:architecture}
  
\end{figure}

\input{tables/layer_sizes}

\section{Network Architectures}
\label{architectures}
\subsection{CNN}
The CNN architecture mostly follows AlexNet~\cite{krizhevsky2012imagenet}, with modifications to allow for class conditional reconstruction as in Qin et al.~\cite{qin2019detecting}. In particular, we substitute the penultimate fully connected layer in AlexNet with logistic regressions on $k$ embedding class vectors, one for each class in the dataset, each of size $m$. We set $m = 16$, which is the same size as the matrix capsules used for reconstruction in the capsule architectures. We also substitute the last layer in AlexNet to linearly transform each of the $k$ embedding vectors into a logit, stacked together before softmax activation, which yields the probabilities for each of the $k$ classes. 
From the penultimate layer consisted of $k$ logistic regressions, we select a single embedding vector from the $k$ class vectors, which is either the vector for the ground truth class during training time, or the predicted class at test time. 

The selected class embedding vector is then fed into the decoder portion of the network. The decoder consists of two hidden fully connected layers (Hidden FC1, Hidden FC2) with hypertangent activation and an output layer (Reconstruction Output Layer) that has the same number of neurons as the number of pixels in the input image. The reconstructed output is compared to the input image, which yields the \emph{reconstruction loss}. The reconstruction loss is compared to a threshold for detection and minimized during model training and adaptive attack optimization. The input images are scaled to side lengths of 224 in order to be consistent with \cite{krizhevsky2012imagenet}.

\subsection{CapsNet Architecture}
We utilize two CapsNet-based architectures for detection. The baseline model is the one proposed by Hinton et al.~\cite{hinton2018matrix}, while the other model includes the two modifications discussed in Section~\ref{modifications}. They both consist of a classifier and a decoder. 

The classifier is identical to Hinton et al.'s~\cite{hinton2018matrix} in both models, and as shown in Figure~\ref{fig:architecture}, it consists of a convolutional layer (Conv), followed by a primary capsule layer (PrimCaps), followed by two convolutional capsule layers (ConvCaps1, ConvCaps2), and finally followed by a class capsule layer (ClassCaps). As in Hinton et al.~\cite{hinton2018matrix}, we have kernel size of 5 in Conv, 1 in PrimCaps, and 3 in ConvCaps1 and ConvCaps2. The numbers of convolutional kernels, capsule convolutional kernels, capsules, and fully connected neurons differ by the dataset (Table~\ref{table: layer_sizes}), but are the same for our two models. 
The class capsule layer outputs a classification loss for model training or an adversarial cost for attack optimization. The classification loss is a modified version of \emph{hinge loss}~\cite{hinton2018matrix}.

In both of our CapsNet models, a capsule is selected from the class capsule layer, which is either the capsule for the ground truth class during training time or, at test time, the most likely capsule. The matrix pose of the selected capsule is flattened into a vector and then fed into the decoder. 
The decoder operates just as the decoder for the CNN.

\input{tables/bs_reg}

For training both our models and our patches, we use a learning rate of 0.003 in ADAM optimizer and two routing iterations for expectation maximization. The final temperature hyperparameter for the modified hinge loss is configured to be 0.01 for all the CapsNet training sessions, consistent with ~\cite{hinton2018matrix}. Batch sizes and regularization hyperparameters used to train the CapsNet architectures on different datasets are contained in Table~\ref{table: bs_reg}. Since CapsNets are very slow to train, we terminate their training sessions after 5 days even if they have not yet converged. For other implementation details, refer to our source code to be released after publication.

\subsection{Improved CapsNet}
\label{modifications}

\begin{figure}
    \centering
    \includegraphics[width=7cm]{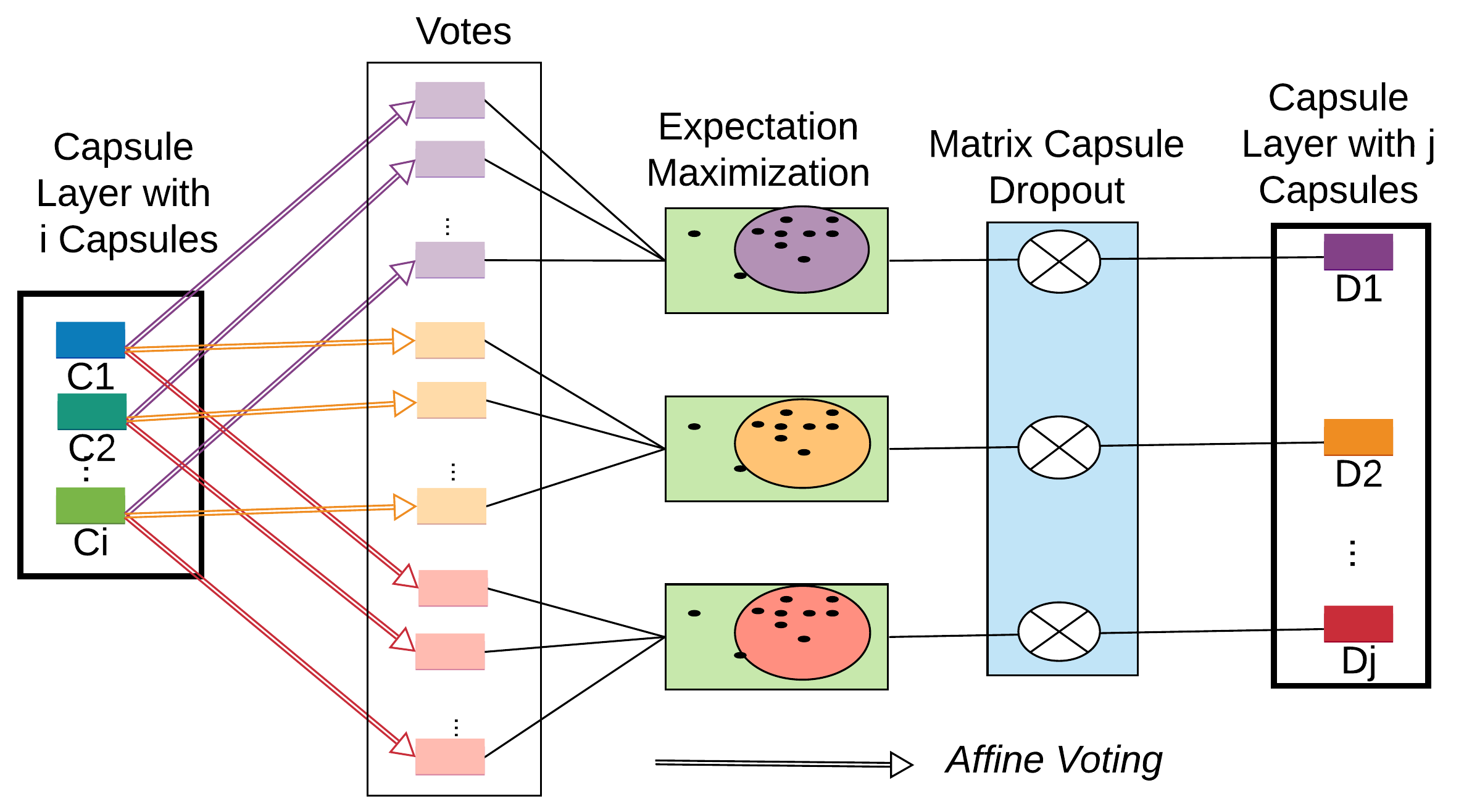}
    \caption{An abstraction of our modified matrix capsule layer, with MCD (Section \ref{mcd}) and affine voting (Section \ref{affine_voting}).} 
    \label{fig:capsule_layer}
\end{figure}

\begin{figure*}
    \centering
    \begin{subfigure}{.5\textwidth}
      \centering
      \includegraphics[scale=0.175]{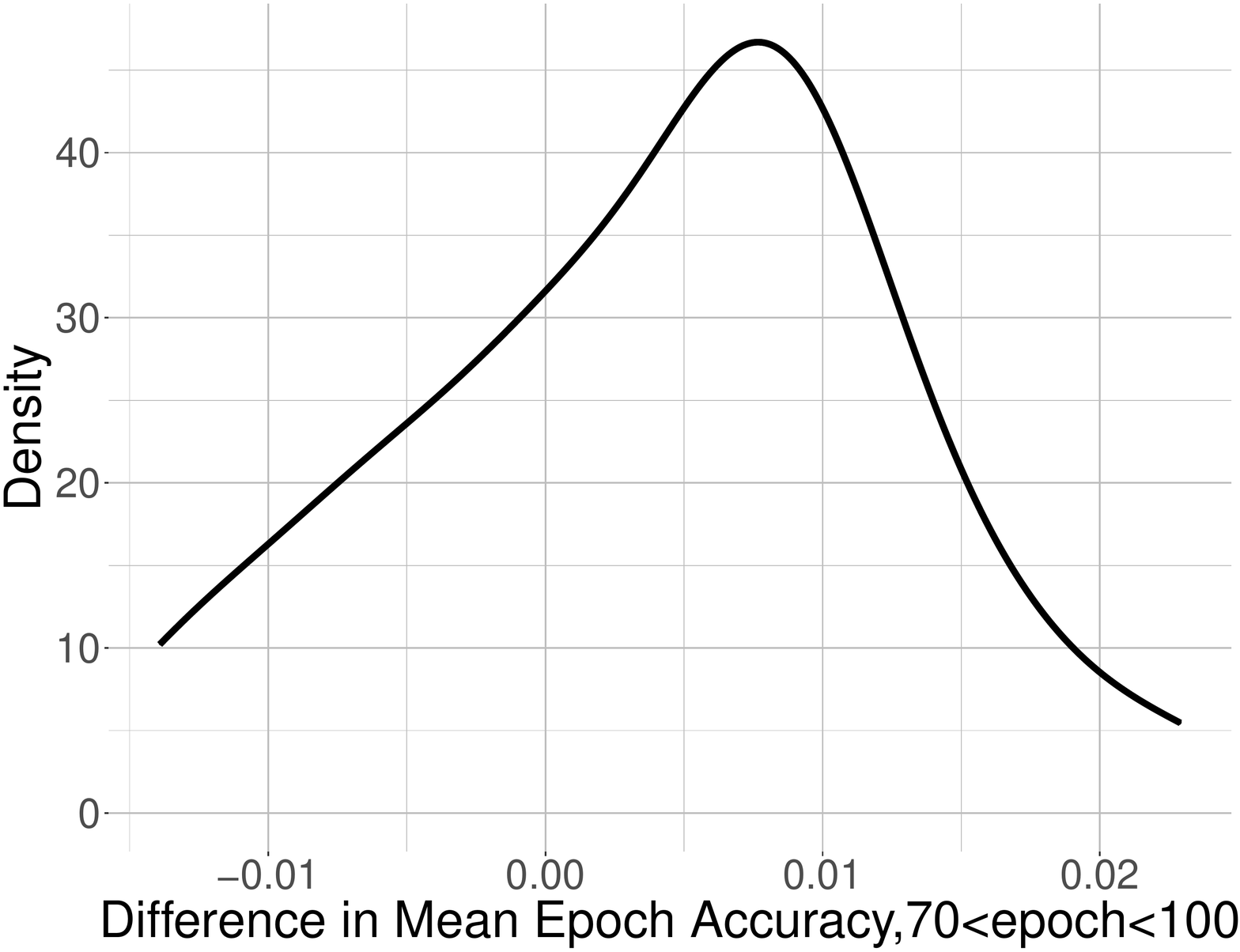}
      \caption{Distribution of differences in test accuracies between 70 and 100 epochs.}
      \label{affine_epochs}
    \end{subfigure}%
    ~
    \begin{subfigure}{.5\textwidth}
      \centering      \includegraphics[scale=0.175]{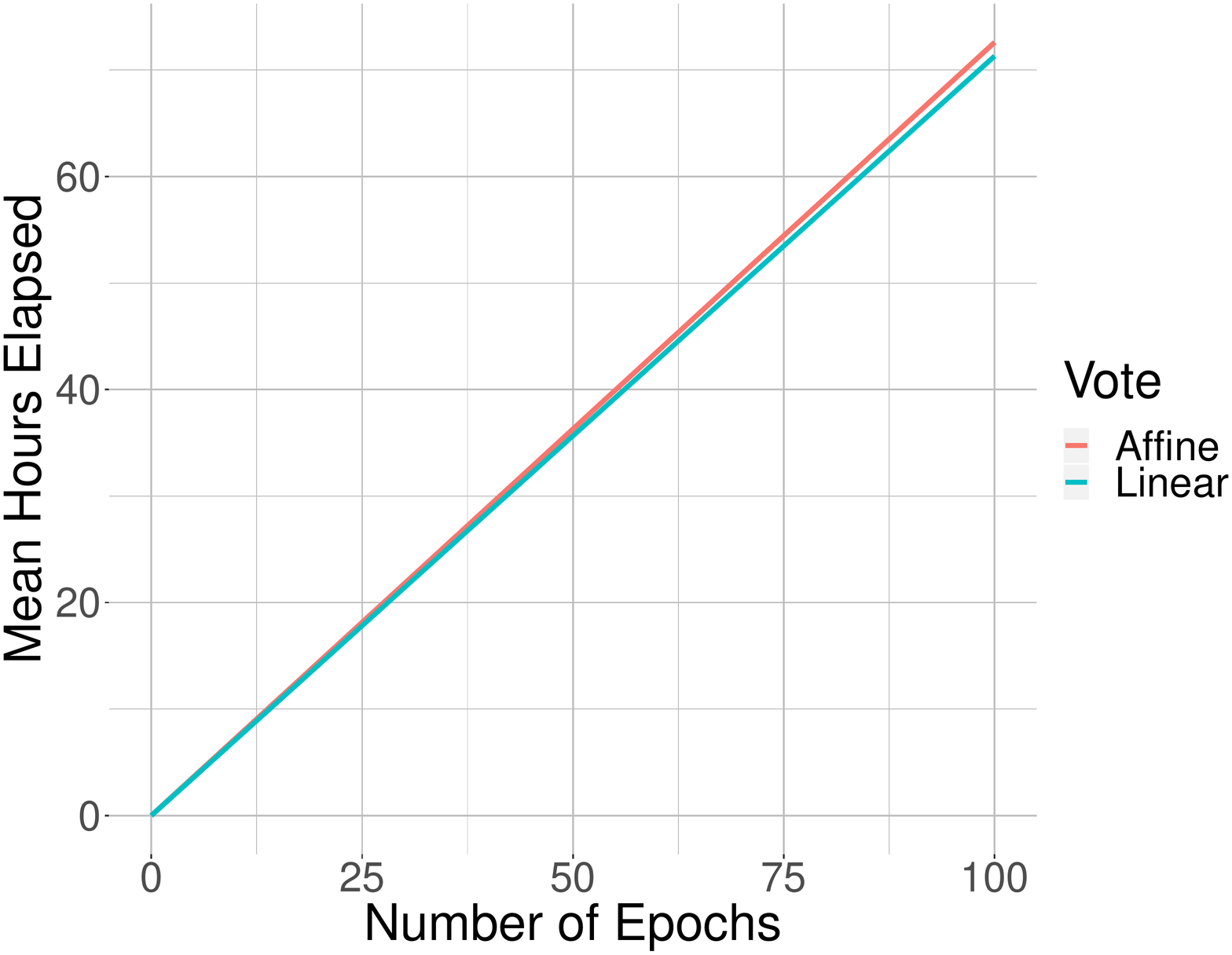}
      \caption{Total training time of the two voting mechanisms.} 
      \label{affine_traintime}
    \end{subfigure}
    \caption{Affine Voting improves post convergence accuracy and has minimal training time impact.}
    \label{fig:affineVSoriginal2}
\end{figure*}

\begin{figure*}[t!]

\centering
    \begin{subfigure}{.5\textwidth}
      \centering
      \includegraphics[scale=0.175]{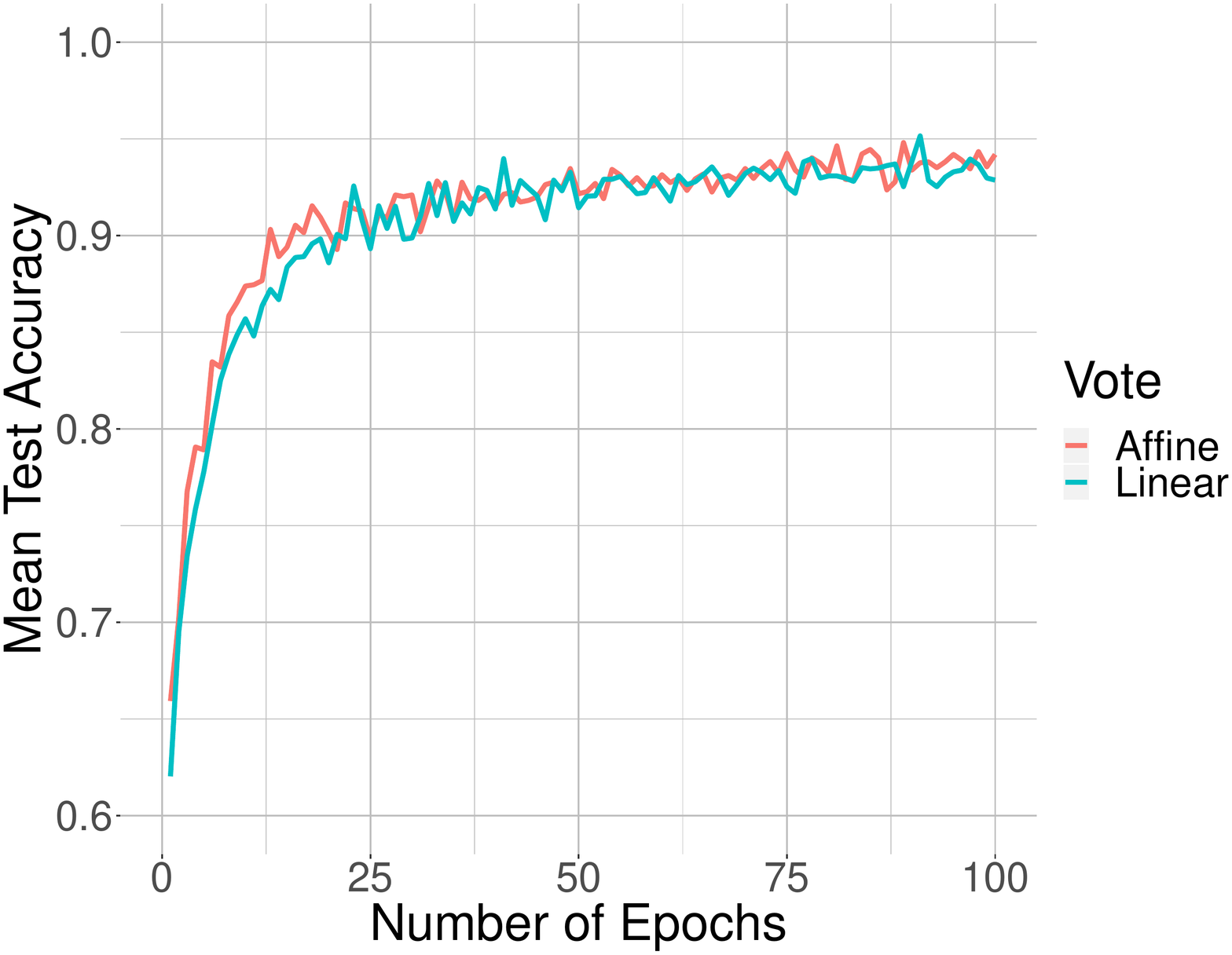}
      \caption{mean test set accuracies over epochs.}
    \end{subfigure}%
    \begin{subfigure}{.5\textwidth}
      \centering
      \includegraphics[scale=0.175]{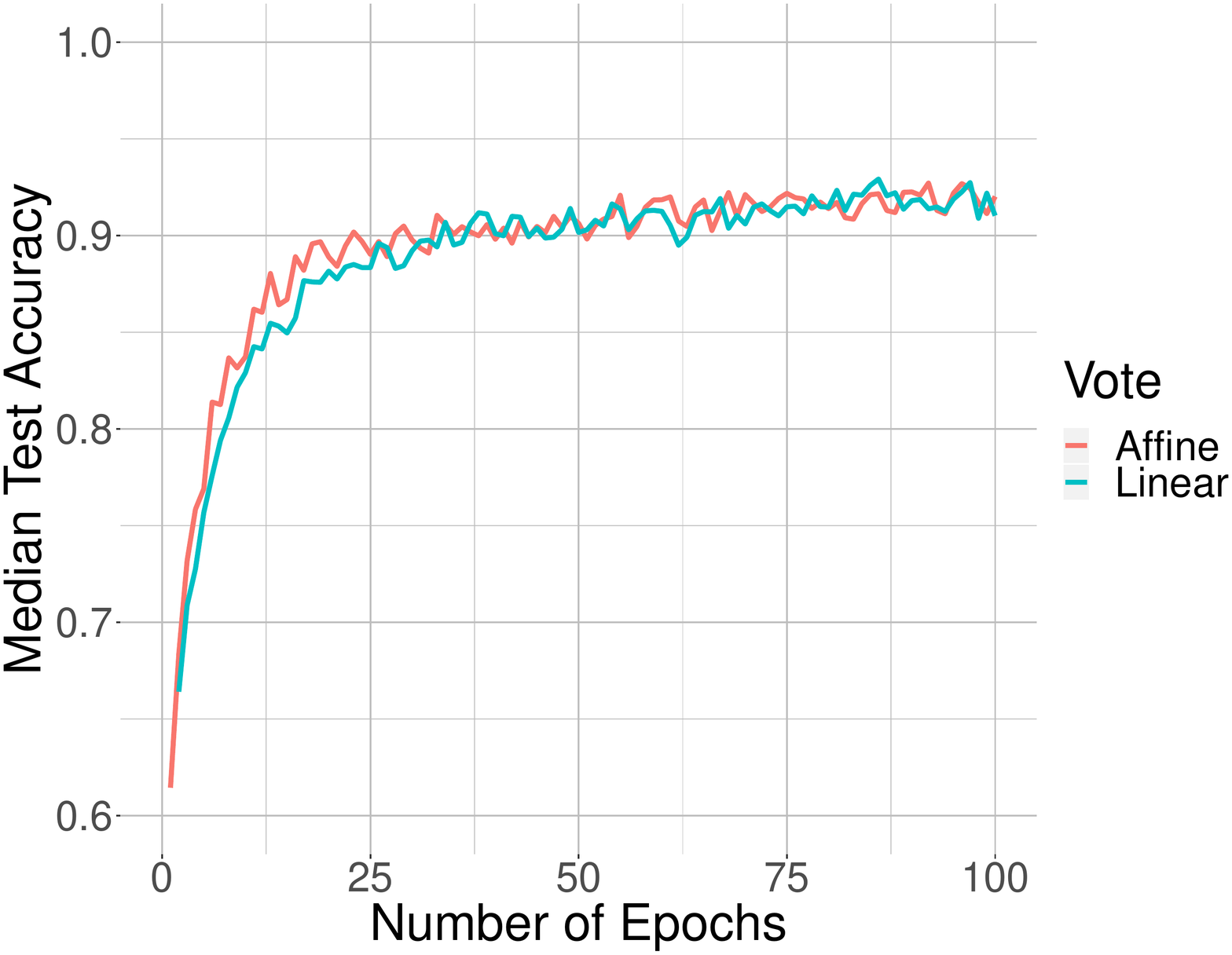}
      \caption{median test accuracies over epochs.}
    \end{subfigure}
\caption{Affine voting speeds up convergence. There is a noticeable gap between training sessions using affine voting and linear voting, roughly between epoch 10 to epoch 25. This shows that affine voting capsules are able to converge faster.}
\label{fig:affineVSoriginal1}
\end{figure*}

We were not able to achieve satisfactory accuracies on SVHN and CIFAR10 using the architecture from Hinton et al.~\cite{hinton2018matrix}. Since inputs that are misclassified are adversarial from the start, with no effort from the attacker, we propose two improvements on the CapsNet architecture to get it on a more equal footing for the comparison with AlexNet. Our Improved CapsNet differs in that the capsule layers are configured with {\em Matrix Capsule Dropout} (MCD) and {\em Affine Voting}.

We provide an intuitive explanation of these improvements along with empirical evidence on the smallNORB dataset. To the best of our knowledge, we are the first to attempt dropout and affine voting on matrix capsule layers (Figure \ref{fig:capsule_layer}).
\input{tables/mcd}
\paragraphX{Matrix-Capsule-Dropout.}
\label{mcd}
{\em Dropout} was proposed by Srivastava et al.~\cite{srivastava2014dropout} in 2014, and since then it has been used in almost all of the state-of-the-art CNN architectures due to its regularizing effects. It approximates ensemble learning in a single dense network by randomly setting neuron activation outputs to zero with probability $p$. We leverage the same intuition in designing {\em matrix-capsule-dropout} (MCD). If a matrix capsule layer has MCD enabled, then the output capsule activation probabilities of its EM routing procedure during forward propagation will be randomly zero-masked with probability $p$ after the last expectation maximization step.

This is a computationally inexpensive procedure, as we are only masking the activation probabilities of the capsules without tampering with their matrix representation. In our model, we set $p=0.5$ and apply MCD to randomize the capsule activation probabilities of the second to last and third to last layers in the classification network, so the hidden capsules voting for class capsules in the last layers are deactivated randomly. We are able to {\em reduce over-fitting and improve test accuracy} with this implementation, as shown in Table~\ref{table:dropout_results}. The choices regarding $p$ and affected layers were made without hyperparameter optimization, so better results may be possible. To the best of our knowledge, this is the first work to implement dropout on matrix capsule layers.

We also attempted {\em DropRoute}, an adaptation of DropConnect~\cite{wan2013regularization} on the Routing by Agreement mechanism, by randomly assigning routing weights to zero after each expectation maximization step. As we can see from Table~\ref{table:dropout_results}, {\em DropRoute} was not effective.
\paragraphX{Affine Voting.}
\label{affine_voting}
Hinton et al. mention that a lower-level capsule calculates its votes for capsules in the higher levels by ``multiplying its own pose matrix by trainable viewpoint-invariant transformation matrices that could learn to represent part-whole relationships''~\cite{hinton2018matrix}. In other words, votes on higher-layer capsules' pose matrices are linear transformations of lower-layer capsules. Geometrically, the relationships between higher- and lower-layer capsule poses estimated by those linear votes are compositions of rotations, reflections, and scaling.

On the other hand, our architecture calculates the votes through {\em affine transformations}, which multiplies a pose matrix with a trainable transformation matrix, followed by addition of a trainable bias matrix. While this is partly inspired by the use of bias terms in CNNs, geometrically affine transformations can model compositions of rotations, reflections, and scaling, followed by translations~\cite{weisstein2004affine}.
Thus, they are strictly more expressive than that of the default linear transformations proposed in the {\em EM Routing} paper, and can model more general relationships between higher and lower level capsules. This, however, comes at the cost of doubling the amount of trainable parameters in capsule layers with all else unchanged.

To evaluate the effect of affine transformations, we compared the training and testing performances of {\em linear voting} and {\em affine voting} on the smallNORB dataset, with ten training sessions for each voting mechanism. We find that affine voting significantly increases post-convergence accuracy. In Figure~\ref{affine_epochs}, we illustrate the empirical density of differences in test accuracies between affine voting and linear voting in the range from 70 and 100 epochs, where the training has converged. The density is normally distributed (Shapiro-Wilk p=0.840), and we find a statistically significant improvement for affine voting, as indicated by a rejection of the hypothesis that the differences are centered at zero (one-tailed T-test, p=0.010). Furthermore, affine voting only increases the time per epoch by about 2\% (Figure~\ref{affine_traintime}),
while speeding up convergence overall by reducing the number of training epochs needed to get good accuracies (Figure~\ref{fig:affineVSoriginal1}). Note that our training times reached over 60 hours for 100 epochs, so the trade-off of a relatively lower accuracy for a faster training time may be worthwhile in some applications.

%% file: tables/layer_sizes.tex
\begin{table}[!t]
    \centering
    \begin{tabular}{l||l|l|l|l|l|l|l|l}
    \hline
    Dataset   & A & B & C & D & E & X & Y\\\hline
    smallNORB & 64 & 8 & 16 & 16 & 5 & 0 & 0 \\
    MNIST     & 64 & 8 & 16 & 16 & 10 & 512 & 1024 \\
    SVHN      & 128 & 32 & 24 & 32 & 10 & 512 & 1024 \\
    CIFAR10   & 128 & 24 & 24 & 32 & 10 & 512 & 1024 \\\hline
    \end{tabular}
    \vspace{0.1cm}
    \caption{Number of convolutional kernels (A - Conv), number of capsule kernels (B - PrimCaps, C - ConvCaps1, D - ConvCaps2), number of class capsules (E - ClassCaps), and number of hidden neurons (X - Hidden FC1, Y - Hidden FC2) for the layers in our architecture on different datasets. Other than the number of class capsules, they were arbitrarily selected for each dataset. For experiments conducted on the same dataset, these numbers remain unchanged, even if other hyperparameters settings such as $\lambda_{a}$ or architectural settings such as MCD may differ.}
    \label{table: layer_sizes}
    \vspace{-0.6cm}
\end{table}

%% file: tables/bs_reg.tex
\begin{table}[!ht]
    \centering
    \begin{tabular}{l||l|l|l|l}
    \hline
    \multirow{2}{*}{\textbf{Dataset}}   & \multirow{2}{*}{\textbf{$\lambda_{capsule}$}} & \multirow{2}{*}{$\lambda_{conv, fc}$} & model  & patch \\ 
    & & & batch size & batch size \\\hline
    smallNORB & 0 & 2e-07 & 64 & N/A\\
    MNIST     & 0 & 2e-07 & 128 & 128\\
    SVHN      & 1e-06 & 1e-06 & 24 & 16\\
    CIFAR10   & 0 & 2e-07 & 16 & 16\\\hline
    \end{tabular}
    \vspace{0.1cm}
    \caption{Regularization parameters and batch sizes for training the CapsNet architecture.}
    \label{table: bs_reg}
    \vspace{-0.4cm}
\end{table}

%% file: tables/mcd.tex
\begin{table*}[ht]
    \centering
    \begin{tabular}{l|l|l|l|l|l|l}
    \hline
    & \multicolumn{2}{c|}{Unmodified} & \multicolumn{2}{c|}{MCD} & \multicolumn{2}{c}{DropRoute} \\ \cline{2-3} \cline{4-5} \cline{6-7}
    & Train Acc.      & Test Acc.
    & Train Acc.      & Test Acc.
    & Train Acc.      & Test Acc.\\ \hline
    
    & 100.00\% & 94.52\%
    & 99.99\%  & 95.52\%
    & 99.82\% & 94.31\%\\
    
    & 99.99\% & 94.04\%
    & 99.79\%  & 95.44\%
    & 99.80\% & 93.69\%        \\
    
    & 99.99\% & 94.48\%
    & 99.96\% & 95.72\%
    & 99.65\% & 94.70\%        \\
    
    & 99.99\% & 94.28\%
    & 100.00\% & 95.52\% 
    & 99.85\% & 93.69\%       \\\hline
    
    {\bf Mean}
    & \textbf{99.99\%} & \textbf{94.33\%}
    & \textbf{99.94\%}  & \textbf{95.55\%}
    & \textbf{99.78\%} & \textbf{94.10\%}      \\ 
    \hline \hline
    \end{tabular}
    \vspace{0.1cm}
    \caption{Statistics from twelve 120-hour training sessions on smallNORB dataset (lasting 220-240 epochs each). The train partition accuracies and test partition accuracies from the epoch checkpoints with the best test partition accuracies are reported, for each training session. We can see that matrix-capsule-dropout (MCD) noticeably reduces over-fitting.}
    \label{table:dropout_results}
    \vspace{-0.7cm}
\end{table*}

%% file: 5_exp_setup.tex
\section{Detection Experiment Setup}
\label{exp_setup}
We now present the setup for our adversarial patch detection experiments. Section~\ref{adv_patching} elaborates on the optimization objective for training the adversarial patches, while Section~\ref{metrics} defines the evaluation metrics. Note that since physical adversarial attacks are generally first created digitally then transferred over to the physical domain~\cite{athalye2017synthesizing, brown2017adversarial}, it is plausible that the effectiveness of the detector on the digital attacks also transfers to the physical domain.

\input{tables/patch_MNIST}

\subsection{Adversarial Optimization Objective}
\label{adv_patching}
We use two types of adversarial patching attacks, a {\em naive whitebox attack} and an {\em adaptive whitebox attack}, to assess detection performance.

Both attacks use an overarching optimization objective based on {\em Expectation over Transformation} (EOT)~\cite{athalye2017synthesizing}. 
With EOT, optimization objective in both attacks can be expressed as:
\begin{gather}
    \theta_{patch} = \argmin_{\theta_{patch}}\E_{x\sim X, t\sim T, l\sim L}[J_{adv}(x, t, l, P, \theta_{patch}, A)], \label{eq:overall_objective}
\end{gather}

where $\theta_{patch} \in R^{w', h', 3}$ are parameters for the patch of width $w'$ and height $h'$, $X$ is the training set of images, $T$ is the distribution of transformations, $L$ is the distribution of locations over images, $J_{adv}$ is the scalar valued adversarial cost function, $P: R^{w', h', 3} \rightarrow [0, 1]^{w', h', 3}$ is a differentiable function transforming the patch parameters into an RGB patch, and $A: [0, 1]^{w', h', 3}, [0, 1]^{w, h, 3}, T, L \rightarrow [0, 1]^{w, h, 3}$ is a differentiable function applying the patch onto the training RGB image with width $w$ and height $h$. Note that the RGB values in this paper are floats between 0 and 1 due to dataset normalization. For both attacks,
\begin{gather}
    P(\theta_{patch}) = (tanh(\theta_{patch}) + 1_{w'\times h'}) / 2
\end{gather}
which transforms the patch parameters in $R^{w', h', 3}$ to $[0, 1]^{w, h, 3}$, and thus prevents the need for clipping functions to restrict the image in normalized RGB space. This improves the effectiveness of gradient descent on the smoother objective function and is inspired by the {\em Change of Variables} approach from~\cite{carlini2017towards}.

Our attacks use the patch applier function $A$ from the source code used by Brown et al.~\cite{brown2017adversarial}, which samples from the distribution of random transformations, locations, training images, and returns a new image with the randomly transformed patch masking a random location on the training image. We also use the same parameters for $A$, except we changed the uniform distribution of patch scaling to be between 10\% and 50\% of the input image instead of between 30\% and 150\%. This was to improve the patching effectiveness at lower scales.

Given the attacker's goal of causing misclassification, the neural network's confidence on the correct class needs to be lower than its confidence on the incorrect classes. Due to the universal nature of physical adversarial objects and patches, the most straightforward method to cause misclassification is to aim at a single adversarial target class $c_{adv}$. In our experiments, we arbitrarily set $c_{adv} = 0$.

Both attacks use an iterative numerical method, namely {\em ADAM}~\cite{kingma2014adam}, to search for the strongest $\theta_{patch}$, due to its relative efficiency compared to other iterative methods and its ability to find stronger adversarial solutions than analytical methods~\cite{carlini2017towards}. It is also relatively insensitive to hyperparameter choices.

The two attacks differ in their adversarial cost function $J_{adv}$, which we define as follows.

\paragraphX{Naive Whitebox Attack.}
For this attack, in complement to the overall optimization framework in Equation~\ref{eq:overall_objective}, we have,
\begin{gather}
    J_{naive}(x, t, l, P, \theta_{patch}, A) 
    = \max(\max_{i\neq c_{adv}}(Z(x')_{i})-Z(x')_{c_{adv}})^{+}, -\kappa) \label{eq:naive_cost}
\end{gather}
where $Z(x')_{i}$ is the logit for the activation of class $i$ capsule, $\kappa$ is the ideal confidence gap between $c_{adv}$ and the next most confident class, and $x'$ is the adversarially patched image. We can define $x'$ as:
\begin{gather}
    x' = A(P(\theta_{patch}), x, t, l) \label{eq:adv_input}
\end{gather}
This adversarial cost function is similar to the most powerful objective identified by Carlini and Wagner~\cite{carlini2017towards}, with the two differences being that we optimize over batches of adversarially patched images rather than perturbation over a single image, and we do not seek to minimize the distance between $x'$ and $x$, since we expect the patch to be visible to human observers. We set $\kappa = 20$, as this confidence gap was found to increase the transferability of adversarial perturbations~\cite{carlini2017towards}. It is possible that other values of $\kappa$ could achieve better adversarial patching results.

\paragraphX{Adaptive Whitebox Attack.}
As we seek to both fool the classifier and evade the detector (Equation~\ref{eq:detection}) with this attack, we modify the adversarial cost function $J_{adv}$ to include the constraint that $f_{detection}(x) = 0$. However, as the Heaviside step function does not have a non-zero gradient outside the decision boundary, the Lagrangian method cannot directly be used for this constraint. Instead, we rephrase the constraint as $\Delta_{recon} = 0$, and include a Lagrangian term for this constraint so that the generated patch will seek to not only fool the detector but also minimize the reconstruction loss. We then have, 
\begin{gather}
    J_{adaptive}(x, t, l, P, \theta_{patch}, A, \lambda_{a}) \notag\\
    = J_{naive}(x, t, l, P, \theta_{patch}, A) +  \lambda_{a} * \Delta_{recon},  \label{eq:adaptive_cost}
\end{gather}
where $\lambda_{a}$ is a new hyperparameter that serves as the Lagrangian multiplier and balances the objective to fool the classifier and the constraint to evade the detector. To determine the strongest $\lambda_{a}$ that yields the highest proportion of undetected successful attacks, we utilized a simple grid search on [0,0.5,1,1.5,2,2.5,3,3.5, 4,4.5,5,10,100,1000,10000]. 

\vskip 0.4cm
In our experiments, patches developed with the two methods are applied to and optimized over the test set images, in contrast to the CapsNet models, which are trained on the training set images. This is to maximize the effectiveness and the threat of adversarial patches on the test set images. We use the same batch sizes and learning rates to train the CapsNet patches as we used to train the CapsNet models in Section~\ref{architectures}. This is also the case for training the CNN patches on MNIST. However, we used a higher learning rate (0.01) to train the CNN patches on SVHN and CIFAR10, in order to reduce underfitting. 

\subsection{Evaluating the Patches}
\label{metrics}
For each dataset, we separately trained the patches against the three architectures (CNN, CapsNet, and improved CapsNet) with different values of $\lambda_{a}$, using the optimization objective from Equation~\ref{eq:adaptive_cost}. Note that when $\lambda_{a} = 0$, the adaptive attack (Equation~\ref{eq:adaptive_cost}) is identical to the naive attack (Equation~\ref{eq:naive_cost}), and evaluating the adaptive attack with $\lambda_{a} = 0$ is as same as evaluating the naive attack. We thus highlight statistics for the $\lambda_{a}$ that results in the highest adversarial success rates (ASR) for each dataset, as well as the statistics for $\lambda_{a} = 0$ to compare naive and adaptive attacks, in Tables~\ref{table:mnist_results}, \ref{table:svhn_results}, and \ref{table:cifar10_results}.

\input{tables/patch_SVHN}

\paragraphX{Adversarial Success Rate (ASR).}
We define ASR as the proportion of adversarial inputs that both fool the classifier and evade the detector on test set images. ASR is an overall metric that can be interpreted as how effective the adversarial patch is against a protected model that has a detector.

\paragraphX{Adversarial Fooling Rate (AFR).} We also show AFR, the proportion of adversarial inputs that fool the classifier. AFR is only impacted by the effectiveness of the patch on the classifier and not the detector. AFR can be interpreted as how effective the adversarial patch is against the model unprotected by a detector. 

\paragraphX{Adversarial Detection Rate (ADR).} We define ADR as the proportion of adversarial inputs that are detected as adversarial. ADR can be interpreted as the recall rate for the detector, the higher it is, the better it is for the detector. 

Since we are using targeted attacks with target class $c_{adv} = 0$, input images with a ground truth of $y = 0$ are not included in the analysis. The detection false positive rate is controlled at 0.05 per Section~\ref{decoder}. 

During evaluation, the patches are applied at random locations and rotations as they were during training, but they are scaled to cover 40\% of the input image instead of a randomly sized area. The reason we use 40\% instead of smaller sizes is that the patches are ineffective at fooling the classifier if scaled any smaller. To demonstrate this, we plot (in the Appendix) the $ADR$, $AFR$, and $ASR$ at different scales for the naive ($\lambda_{a} = 0$) patch and most effective adaptive ($\lambda_{a} \neq 0$) patch for each detector on each dataset. We also show example patches in the Appendix as well.

%% file: tables/patch_MNIST.tex
\begin{table*}[t]
    \centering
    \begin{tabular}{|l|l|l|l|l|l|l|l|l|l|}
    \hline 
    & \multicolumn{3}{c|}{CNN} & \multicolumn{3}{c|}{CapsNet} & \multicolumn{3}{c|}{Improved CapsNet} \\ 
    & \multicolumn{3}{c|}{Test Acc.: 99.50\%} & \multicolumn{3}{c|}{Test Acc.: 99.32\%} & \multicolumn{3}{c|}{Test Acc.: 99.50\%} \\ \cline{2-4} \cline{5-7} \cline{8-10}
    $\lambda_{a}$ & ADR      & AFR      & ASR      & ADR     & AFR      & ASR      & ADR      & AFR      & ASR      \\ \hline
    0             
    & \textbf{0.180235} &  \textbf{0.002551} &  \textbf{0.000887}      
    &  \textbf{0.985786} &  \textbf{0.052193} &  \textbf{0.002554}         
    &  \textbf{0.987896} &  \textbf{0.026430} &  \textbf{0.001222}         \\
    
    0.5
    & 0.167258 &  0.002329 &  0.001331          
    & 0.962798 &  0.042199 &  0.003665
    &  0.971793 &  0.026874 &  0.004664        \\
    
    1  
    & 0.194099 &  0.001664 &  0.000998        
    &  \textbf{0.869961} &  \textbf{0.025208} &  \textbf{0.005552}    
    &  \textbf{0.812660} &  \textbf{0.017324} &  \textbf{0.005997}         \\
    
    1.5          
    & 0.160825 &  0.002329 &  0.001664        
    &  0.748917 &  0.015547 &  0.004220  
    &  0.604886 &  0.010661 &  0.003443 \\
    
    2  
    & 0.174579 &  0.001996 &  0.001331        
    &  0.605775 &  0.010105 &  0.002443    
    &  0.522710 &  0.010883 &  0.003887         \\
    
    2.5
    & 0.190772 &  0.001775 &  0.000998           
    &  0.514936 &  0.008773 &  0.002332     
    &  0.472959 &  0.008884 &  0.003554         \\
    
    3  
    & 0.154281 &  0.001886 &  0.001442      
    &  0.459856 &  0.007885 &  0.001444   
    &  0.443753 &  0.006996 &  0.002443         \\
    
    3.5
    & 0.165373 &  0.002773 &  0.001775         
    &  0.446086 &  0.008440 &  0.001333    
    &  0.421877 &  0.007440 &  0.003554         \\
    
    4
    & 0.166260 &  0.001442 &  0.000887
    &  0.440422 &  0.007551 &  0.001888
    &  0.407773 &  0.006108 &  0.003220         \\
    
    4.5
    & 0.164485 &  0.002884 &  0.001886           
    &  0.435980 &  0.008329 &  0.002110
    &  0.379678 &  0.005441 &  0.002665         \\
    
    5  
    & 0.165594 &  0.002329 &  0.001886         
    &  0.429872 &  0.007885 &  0.001110
    &  0.363021 &  0.004109 &  0.001999          \\
    
    10 
    & 0.158829 &  0.001109 &  0.000776        
    &  0.420988 &  0.005886 &  0.001333
    &  0.356469 &  0.003665 &  0.001777         \\
    
    100
    & 0.082298 &  0.002107 &  0.001775           
    & 0.423431 &  0.006996 &  0.000777
    &  0.360355 &  0.005108 &  0.002443         \\
    
    1000          
    &  0.074091 &  0.001886 &  0.001553        
    &  0.429539 &  0.007218 &  0.000888
    & 0.359134 &  0.003554 &  0.001777         \\
    
    10000         
    & \textbf{0.077751} &  \textbf{0.002551} &  \textbf{0.002218}          
    &  0.423987 &  0.007218 &  0.001110 
    & 0.359578 &  0.003554 &  0.001333        \\
    \hline 
    \end{tabular}
    \vspace{0.2cm}
    \caption{Evaluation statistics of patches on the MNIST dataset, randomly applied to 40\% of input image area. Statistics for the {\em naive attack ($\lambda_{a} = 0$)} and the {\em adaptive attack ($\lambda_{a} > 0$)} with the best $ASR$ are in bold text.
    }
    \label{table:mnist_results}
\end{table*}

%% file: tables/patch_SVHN.tex
\begin{table*}[th]
    \centering
    \begin{tabular}{|l|l|l|l|l|l|l|l|l|l|}
    \hline 
    & \multicolumn{3}{c|}{CNN} & \multicolumn{3}{c|}{CapsNet} & \multicolumn{3}{c|}{Improved CapsNet} \\ 
    & \multicolumn{3}{c|}{Test Acc.: 92.42\%} & \multicolumn{3}{c|}{Test Acc.: 90.92\%} & \multicolumn{3}{c|}{Test Acc.: 94.8\%} \\ \cline{2-4} \cline{5-7} \cline{8-10}
    $\lambda_{a}$ & ADR      & AFR      & ASR      & ADR     & AFR      & ASR      & ADR      & AFR      & ASR      \\ \hline
    0             
    &  \textbf{0.248548} &  \textbf{0.026148} &  \textbf{0.019354}         
    &  \textbf{0.835289} &  \textbf{0.147251} &  \textbf{0.022895}         
    &  \textbf{0.606678} &  \textbf{0.161973} &  \textbf{0.065217}         \\
    
    0.5
    & \textbf{0.301668} &  \textbf{0.034713} &  \textbf{0.025983}          
    &  0.854478 &  0.155528 &  0.025860
    &  0.628376 &  0.167078 &  0.058753         \\
    
    1  
    &  0.253037 &  0.025942 &  0.019559          
    &  0.767181 &  0.156805 &  0.044472    
    &  0.534956 &  0.162055 &  0.076910         \\
    
    1.5          
    &  0.253572 &  0.025077 &  0.019106       
    &  0.467861 &  0.149146 &  0.083467  
    &  0.531538 &  0.156868 &  0.072587 \\
    
    2  
    &  0.257031 &  0.022112 &  0.017089          
    &  0.142598 &  0.065596 &  0.058596   
    &  0.498600 &  0.149457 &  0.073164         \\
    
    2.5
    &  0.292938 &  0.032777 &  0.024707    
    &  \textbf{0.175046} &  \textbf{0.121762} &  \textbf{0.105950}     
    &  0.502470 &  0.172843 &  0.087080         \\
    
    3  
    &  0.290097 &  0.033560 &  0.025654         
    &  0.151081 &  0.110397 &  0.098744     
    &  0.435359 &  0.162097 &  0.089962         \\
    
    3.5
    &  0.236813 &  0.020136 &  0.014577         
    &  0.107227 &  0.061561 &  0.057113   
    &  0.410985 &  0.150856 &  0.090909         \\
    
    4
    &  0.252954 &  0.021289 &  0.016636
    &  0.110397 &  0.079020 &  0.073461
    &  0.368454 &  0.144928 &  0.093462         \\
    
    4.5
    &  0.243031 &  0.025489 &  0.019106           
    &  0.110644 &  0.082232 &  0.076343
    &  0.378047 &  0.152297 &  0.095850         \\
    
    5  
    &  0.242619 &  0.020877 &  0.016306          
    &  0.106280 &  0.077208 &  0.072102
    &  \textbf{0.330781} &  \textbf{0.141881} &  \textbf{0.097908}         \\
    
    10 
    &  0.247643 &  0.030389 &  0.023760          
    &  0.083385 &  0.049043 &  0.046613 
    &  0.081316 &  0.063282 &  0.060647         \\
    
    100
    &  0.128227 &  0.027754 &  0.024501           
    &  0.066337 &  0.025530 &  0.023965
    &  0.056448 &  0.005517 &  0.005147         \\
    
    1000          
    &  0.082438 &  0.018571 &  0.016348          
    &  0.065761 &  0.023060 &  0.021618
    &  0.056777 &  0.005147 &  0.004694         \\
    
    10000         
    &  0.079185 &  0.014906 &  0.013012          
    &  0.066543 &  0.022483 &  0.021124 
    &  0.056283 &  0.004488 &  0.004076         \\
    \hline 
    \end{tabular}
    \vspace{0.2cm}
    \caption{Evaluation statistics of patches on the SVHN dataset, randomly applied to 40\% of input image area. Statistics for the {\em naive attack ($\lambda_{a} = 0$)} and the {\em adaptive attack ($\lambda_{a} > 0$)} with the best $ASR$ are in bold text.
    }
    \label{table:svhn_results}
    \vspace{-0.8cm}
\end{table*}

%% file: 6_exp_results.tex
\section{Experimental Results and Analysis}
\label{adv_experiments}

In this section, we discuss the experimental results of adversarial detection on MNIST, SVHN, and CIFAR10 datasets, using the CNN and both the original and improved CapsNet. The effectiveness of adversarial patches against the classifier and the detector are displayed in Table~\ref{table:mnist_results}, Table~\ref{table:svhn_results} and Table~\ref{table:cifar10_results} for MNIST, SVHN, and CIFAR10 datasets, respectively. We will only analyze the bolded statistics, which correspond to the naive attack and the best adaptive attack for an architecture on a dataset, unless otherwise stated. 


\subsection{Results}
\paragraphX{Results on MNIST.}
We first observe that the overall fooling rates ($AFR$s) are quite low across the board at between 0.2-5.2\%, leading to low attacker success rates ($ASR$) of between 0.1-0.6\%. MNIST's low resolution makes it difficult to craft successful attacks. In terms of detection ($ADR$), both CapsNets substantially outperform the CNN at detecting the attack, with over 81\% $ADR$ compared with just 7.8\% $ADR$ for the best adaptive attack. The improved CapsNet is slightly worse overall than the original CapsNet on both $ADR$ and $ASR$, showing that the original CapsNet is better in detecting attacks, though it features lower $AFR$.

\paragraphX{Results on SVHN.}
First, we note that our improved CapsNet outperforms CNN and the original CapsNet on benign test accuracy by more than 2\% and 3.5\%, respectively, showing the benefits of our modifications. Against both naive and adaptive attacks, however, the CNN offers very low fooling rates and thereby performs the best with just 1.9\% $ASR$ for naive attacks and 2.6\% $ASR$ for the best adaptive attack. The improved CapsNet has a significantly worse detection rate ($ADR$) than the original CapsNet, but it shows improvements against adaptive attacks with 33.1\% $ADR$ versus 17.5\% $ADR$ for original CapsNet.

\paragraphX{Results on CIFAR10.}
On CIFAR10, the CNN offers the best benign test accuracy at 80.1\%, 13.8\% better than CapsNet and 4.6\% better than the improved CapsNet. The fooling rate ($AFR$) is much higher on CIFAR10 than the previous two datasets at between 30-57\% for the settings with the best $ASR$ results. Here, the naive attack is best countered by the CNN at 23.5\% $ASR$, but the adaptive attack is best countered by the improved CapsNet at 29.4\% $ASR$, which is 6.2\% and 17.8\% better than the CNN and original CapsNet, respectively. The improved CapsNet's performance on adaptive attacks is marked by both the highest detection rate and the lowest fooling rate of the three models.

\input{tables/patch_CIFAR10}

\subsection{Analysis}
\paragraphX{Does the Detection Work?}
First, we see that in every set of bolded statistics, including the adaptive patches, $ASR$ is lower than $AFR$ by a noticeable amount. This means that a smaller proportion of patched images both fool the classifier and evade detector, than fooling the classifier alone. In other words, some otherwise successful attacks are being detected by the architecture, for every naive attack and best adaptive attack on each dataset with each architecture. Had the detector been ineffective, we would see similar numbers for $ASR$ and $AFR$.

This is most pronounced on the CapsNet architectures on MNIST (Table~\ref{table:mnist_results}), where for the improved CapsNet, the best adaptive attack has $ASR \approx 0.005$ and $AFR \approx 0.025$, which means only around a fifth of the attacks that cause misclassification are undetected. Even on CIFAR10 (Table~\ref{table:cifar10_results}), where the difference is less significant, we have $ASR \approx 0.79AFR$ on the CNN architecture. This shows that the detector retains some of its potency even against adaptive patch attacks, which is in stark contrast with the results from Qin et al.~\cite{qin2019detecting} on adversarial example detection. in which the detectors typically lose their effectiveness against an adaptive attack. Albeit not conclusively, this supports our intuition that digital adversarial example detection methods can be transferred to the adversarial patching domain, and that adversarial patches are easier to detect than digital adversarial examples.

\paragraphX{Is the Adaptive Attack Doing Its Job?}
The "aggressiveness" of the adaptive attack is governed by the multiplier $\lambda_{a}$ in Equation \ref{eq:adaptive_cost}. The higher $\lambda_{a}$, the more our adversarial objective is dedicated to minimizing the reconstruction loss, and hypothetically the less the patches become detected. Our data follows the hypothesis, showing that $ADR$ generally decreases as $\lambda_{a}$ increases. Furthermore, on all datasets and against all architectures, the best adaptive patch has a higher $ASR$ than the naive patch. Thus we believe that our adaptive attack is a reasonable attempt.

\paragraphX{Unmodified CapsNet vs. Improved CapsNet.}
The hyperparameters and architecture of CapsNet and improved CapsNet are identical on each dataset, besides the modifications on the improved CapsNet discussed in Section \ref{architectures}. They were manually tuned to improve their test set accuracies on each dataset. Note that the improved CapsNet architectures are able to achieve significantly higher test accuracy (94.8\% on SVHN, 76.2\% on CIFAR10) compared to the unmodified CapsNet (90.92\% on SVHN, 67.99\% on CIFAR10) on the CIFAR10 dataset. This helps make the classifier more robust to adversarial attacks, as fewer input images would be adversarial by default, which explains the huge differences between $ASR$ on the CIFAR10 dataset, not just on the highlighted patches but across all of them. 


\paragraphX{Comparison among Datasets.} 
We can observe from Table~\ref{table:mnist_results},~\ref{table:svhn_results} and~\ref{table:cifar10_results} that $ASR$ across all the three architectures increases with the increase in dataset complexity. $ASR$ on MNIST dataset is less than $0.01$ whereas it is between $0.01$ and $0.1$ for SHVN, and between $0.23$ and $0.47$ for CIFAR10. $AFR$ shows a similar trend in terms of higher data complexity and higher fooling rate, with $AFR < 0.06$ on MNIST, $0.02 < AFR < 0.17$ on SVHN, and $0.30 < AFR < 0.57$ on CIFAR10. $ADR$ also follows the same pattern, the higher the data complexity, $ADR$ tends to be lower in both unmodified CapsNet and improved CapsNet, and in CNN for SVHN and CIFAR10. These results suggest that adversarial patch attacks might behave similarly with adversarial examples in that they are harder to detect on more complex dataset.

MNIST on CNN has a surprisingly low $ADR$ of $0.08$ against the best performing adaptive patch, and $0.18$ against the naive patch. Two possible factors might have contributed to this anomaly. One is that we upscale the input images for the CNN on all datasets to $[224, 244, 3]$ to be consistent with \cite{krizhevsky2012imagenet}, which causes it to perform worse than on the $[28, 28, 3]$ MNIST dataset, compared to the $[32, 32, 3]$ SVHN and CIFAR10 datasets. Another possibility is that since we use the same CNN architecture hyperparameters for all three datasets, the hyperparameters just happen to perform similarly in terms of $ADR$ on the the $[32, 32, 3]$ datasets, and worse on the $[28, 28, 3]$ MNIST dataset. 

\section{Discussion} 

Our detection approach only requires one forward pass without the time and memory requirements of back propagation and is faster than that of SentiNet~\cite{chou2018sentinet}. This allows safety-critical systems to rapidly sound the alarm when an attack is taking place. Nevertheless, for our approach to be used in real world computer vision systems, reconstruction networks needs to be able to reliably scale up to more complex datasets such as ImageNet. On the bright side, we showed that a detection method originally proposed against digital adversarial examples can be effective against adversarial patches with some modifications, and other perturbation detection methods are worth a try to mitigate risks brought on by physical adversarial attacks. 

We also find that the distribution of reconstruction errors tends have higher means with larger variance on CIFAR10 and SVHN than on MNIST. This could be due to the fact that CIFAR10 and SVHN are non-segmented images containing background noise or even multiple classes of objects. If this is indeed the case, then the architecture needs to learn to segment the background noise from the class objects to more effectively detect adversarial attacks on more complex datasets. This might be achievable through new architectures that can learn segmentation in an unsupervised manner, or supervised training on image segmentation. We leave this investigation as a part of future work.

%% file: tables/patch_CIFAR10.tex
\begin{table*}[th]
    \centering
    \begin{tabular}{|l|l|l|l|l|l|l|l|l|l|}
    \hline 
    & \multicolumn{3}{c|}{CNN} & \multicolumn{3}{c|}{CapsNet} & \multicolumn{3}{c|}{Improved CapsNet} \\ 
    & \multicolumn{3}{c|}{Test Acc.: 80.81\%} & \multicolumn{3}{c|}{Test Acc.: 67.99\%} & \multicolumn{3}{c|}{Test Acc.: 76.2\%} \\ \cline{2-4} \cline{5-7} \cline{8-10}
    $\lambda_{a}$ & ADR      & AFR      & ASR      & ADR     & AFR      & ASR      & ADR      & AFR      & ASR      \\ \hline
    0             
    &  \textbf{0.308582} &  \textbf{0.307248} &  \textbf{0.234549}         
    &  \textbf{0.554024} &  \textbf{0.564695} &  \textbf{0.270565}         
    &  \textbf{0.404333} &  \textbf{0.380444} &  \textbf{0.247778}         \\
    
    0.5
    &  0.308359 &  0.304357 &  0.220542           
    &  0.569253 &  0.513673 &  0.229658
    &  0.421778 &  0.352000 &  0.224333         \\
    
    1  
    &  0.278235 &  0.446532 &  0.342152           
    &  0.542019 &  0.587594 &  0.282125    
    &  0.380444 &  0.364222 &  0.246778         \\
    
    1.5          
    &  0.315362 &  0.424522 &  0.310805        
    &  0.469320 &  0.655069 &  0.373055
    &  0.373444 &  0.353000 &  0.238000 \\
    
    2  
    &  0.307248 &  0.350823 &  0.257225          
    &  0.475767 &  0.517119 &  0.282237    
    &  0.345778 &  0.365000 &  0.259333         \\
    
    2.5
    &  0.301356 &  0.467652 &  0.349378           
    &  0.439084 &  0.670187 &  0.397510     
    &  0.333222 &  0.365778 &  0.267000         \\
    
    3  
    &  0.269787 &  0.413851 &  0.317919      
    &  0.407514 &  0.601823 &  0.375500     
    &  0.301667 &  0.317333 &  0.242222         \\
    
    3.5
    &  0.310805 &  0.394842 &  0.294020          
    &  0.400622 &  0.566363 &  0.354936   
    &  0.289667 &  0.248222 &  0.193667         \\
    
    4
    &  0.294131 &  0.386060 &  0.291018
    &  0.371276 &  0.672743 &  0.443753
    &  0.284889 &  0.262222 &  0.203333        \\
    
    4.5
    &  0.316696 &  0.393953 &  0.284349           
    &  0.361272 &  0.585038 &  0.384838
    &  0.238000 &  0.238778 &  0.193667         \\
    
    5  
    &  0.293241 &  0.391285 &  0.295798          
    &  0.335816 &  0.630058 &  0.433859
    &  \textbf{0.252889} &  \textbf{0.365000} &  \textbf{0.294111}         \\
    
    10 
    &  0.317363 &  0.413517 &  0.306025          
    &  \textbf{0.187528} &  \textbf{0.568253} &  \textbf{0.472099}
    &  0.158111 &  0.287889 &  0.253556         \\
    
    100
    &  \textbf{0.243775} &  \textbf{0.448755} &  \textbf{0.355825}           
    &  0.040907 &  0.057803 &  0.055024
    &  0.037889 &  0.024667 &  0.023444         \\
    
    1000          
    &  0.078257 &  0.336483 &  0.316807          
    &  0.040907 &  0.049133 &  0.046354
    &  0.037333 &  0.019000 &  0.017667         \\
    
    10000         
    &  0.048911 &  0.021787 &  0.020787          
    &  0.041018 &  0.047577 &  0.044242
    &  0.039222 &  0.017000 &  0.016000         \\
    \hline 
    \end{tabular}
    \vspace{0.2cm}
    \caption{Evaluation statistics of patches on the CIFAR10 dataset, randomly applied to 40\% of input image area. Statistics for the {\em naive attack ($\lambda_{a} = 0$)} and the {\em adaptive attack ($\lambda_{a} > 0$)} with the best $ASR$ are in bold text.}
    \label{table:cifar10_results}

    \vspace{-0.8cm}
\end{table*}

%% file: 7_conclusion.tex
\section{Conclusion}
We have shown that capsule networks can be improved using matrix-capsule-dropout and affine voting, techniques inspired by convolutional neural network, and that adversarial digital perturbation detection techniques can be used against adversarial physical payloads with some success. This is promising for the small but growing literature on capsule networks and defenses against adversarial physical attacks, as advances in adjacent but more substantial fields (convolutional neural networks and adversarial digital perturbations, respectively) may be transferred over.

%% file: 99_patches.tex

\section{Patches and Statistics Visualization}
\label{patches}
This section contains visualization of patches and their adversarial statistics, by the architecture and dataset they were optimized on.

\begin{figure*}[htb]
    \begin{subfigure}{.45\textwidth}
    \centering
      \centering
      \includegraphics[scale=0.65]{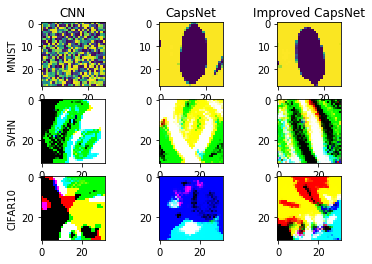}
      \caption{\textbf{Patches.}}
      \label{fig:naive_patches}
      \end{subfigure}
    \begin{subfigure}{.5\textwidth}
      \centering
      \includegraphics[scale=0.45]{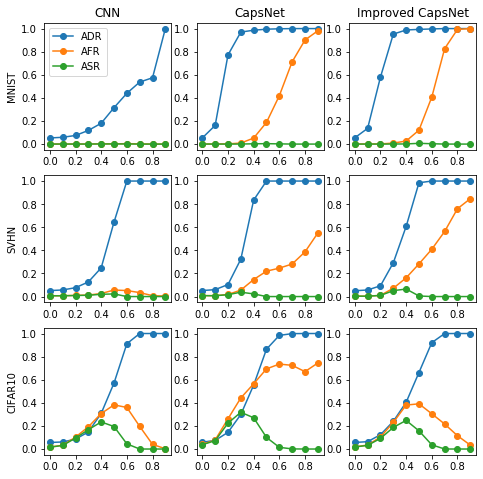}
      \caption{ADR, AFR, and ASR for Patches Scaled to Different Sizes of Input.}
      \label{fig:naive_stats}
    \end{subfigure}
    \caption{Naive Adversarial Attack.}
\end{figure*}

\begin{figure*}[ht]
    \begin{subfigure}{.45\textwidth}
      \centering      \includegraphics[scale=0.65]{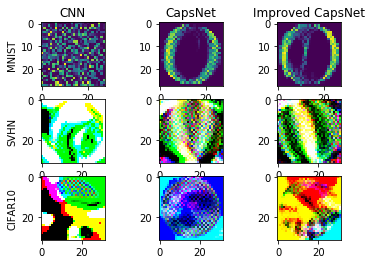}
      \caption{\textbf{Patches.}} 
      \label{fig:adaptive_patches}
    \label{fig:patches}
    \end{subfigure}
    \begin{subfigure}{.5\textwidth}
          \centering
          \includegraphics[scale=0.45]{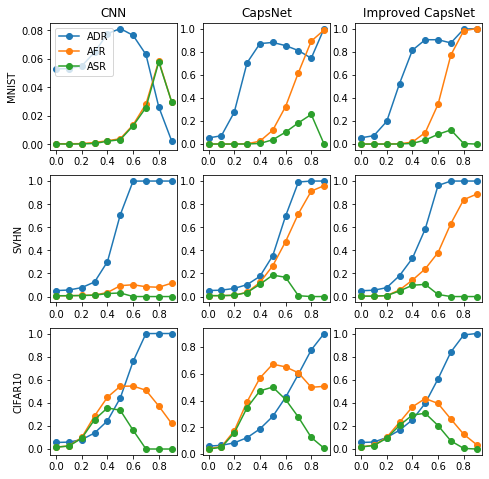}
          \caption{ADR, AFR, and ASR for Patches Scaled to Different Sizes of Input.}
          \label{fig:naive_stats}
    \end{subfigure}    
    \caption{Best Adaptive Adversarial Attack.}
    
\end{figure*}

%% file: main.bbl

\begin{thebibliography}{40}


\ifx \showCODEN    \undefined \def \showCODEN     #1{\unskip}     \fi
\ifx \showDOI      \undefined \def \showDOI       #1{#1}\fi
\ifx \showISBNx    \undefined \def \showISBNx     #1{\unskip}     \fi
\ifx \showISBNxiii \undefined \def \showISBNxiii  #1{\unskip}     \fi
\ifx \showISSN     \undefined \def \showISSN      #1{\unskip}     \fi
\ifx \showLCCN     \undefined \def \showLCCN      #1{\unskip}     \fi
\ifx \shownote     \undefined \def \shownote      #1{#1}          \fi
\ifx \showarticletitle \undefined \def \showarticletitle #1{#1}   \fi
\ifx \showURL      \undefined \def \showURL       {\relax}        \fi
\providecommand\bibfield[2]{#2}
\providecommand\bibinfo[2]{#2}
\providecommand\natexlab[1]{#1}
\providecommand\showeprint[2][]{arXiv:#2}

\bibitem[\protect\citeauthoryear{Athalye, Engstrom, Ilyas, and Kwok}{Athalye
  et~al\mbox{.}}{2017}]%
        {athalye2017synthesizing}
\bibfield{author}{\bibinfo{person}{Anish Athalye}, \bibinfo{person}{Logan
  Engstrom}, \bibinfo{person}{Andrew Ilyas}, {and} \bibinfo{person}{Kevin
  Kwok}.} \bibinfo{year}{2017}\natexlab{}.
\newblock \showarticletitle{Synthesizing robust adversarial examples}.
\newblock \bibinfo{journal}{\emph{arXiv:1707.07397}} (\bibinfo{year}{2017}).
\newblock


\bibitem[\protect\citeauthoryear{Biggio, Corona, Maiorca, Nelson,
  {\v{S}}rndi{\'c}, Laskov, Giacinto, and Roli}{Biggio et~al\mbox{.}}{2013}]%
        {biggio2013evasion}
\bibfield{author}{\bibinfo{person}{Battista Biggio}, \bibinfo{person}{Igino
  Corona}, \bibinfo{person}{Davide Maiorca}, \bibinfo{person}{Blaine Nelson},
  \bibinfo{person}{Nedim {\v{S}}rndi{\'c}}, \bibinfo{person}{Pavel Laskov},
  \bibinfo{person}{Giorgio Giacinto}, {and} \bibinfo{person}{Fabio Roli}.}
  \bibinfo{year}{2013}\natexlab{}.
\newblock \showarticletitle{Evasion attacks against machine learning at test
  time}. In \bibinfo{booktitle}{\emph{Joint European Conference on Machine
  Learning and Knowledge Discovery in Databases}}.
\newblock


\bibitem[\protect\citeauthoryear{Brown, Mané, Roy, Abadi, and Gilmer}{Brown
  et~al\mbox{.}}{2017}]%
        {brown2017adversarial}
\bibfield{author}{\bibinfo{person}{Tom~B. Brown}, \bibinfo{person}{Dandelion
  Mané}, \bibinfo{person}{Aurko Roy}, \bibinfo{person}{Martín Abadi}, {and}
  \bibinfo{person}{Justin Gilmer}.} \bibinfo{year}{2017}\natexlab{}.
\newblock \showarticletitle{Adversarial Patch}.
\newblock \bibinfo{journal}{\emph{arXiv: 1712.09665}} (\bibinfo{year}{2017}).
\newblock


\bibitem[\protect\citeauthoryear{Carlini and Wagner}{Carlini and
  Wagner}{2017a}]%
        {carlini2017bypass}
\bibfield{author}{\bibinfo{person}{Nicholas Carlini} {and}
  \bibinfo{person}{David Wagner}.} \bibinfo{year}{2017}\natexlab{a}.
\newblock \showarticletitle{Adversarial examples are not easily detected:
  Bypassing ten detection methods}. In \bibinfo{booktitle}{\emph{ACM Workshop
  on Artificial Intelligence and Security}}.
\newblock


\bibitem[\protect\citeauthoryear{Carlini and Wagner}{Carlini and
  Wagner}{2017b}]%
        {carlini2017magnet}
\bibfield{author}{\bibinfo{person}{Nicholas Carlini} {and}
  \bibinfo{person}{David Wagner}.} \bibinfo{year}{2017}\natexlab{b}.
\newblock \bibinfo{title}{{MagNet} and "Efficient Defenses Against Adversarial
  Attacks" are Not Robust to Adversarial Examples}.
\newblock
\newblock


\bibitem[\protect\citeauthoryear{Carlini and Wagner}{Carlini and
  Wagner}{2017c}]%
        {carlini2017towards}
\bibfield{author}{\bibinfo{person}{Nicholas Carlini} {and}
  \bibinfo{person}{David Wagner}.} \bibinfo{year}{2017}\natexlab{c}.
\newblock \showarticletitle{Towards evaluating the robustness of neural
  networks}. In \bibinfo{booktitle}{\emph{IEEE symposium on security and
  privacy (S\&P)}}.
\newblock


\bibitem[\protect\citeauthoryear{Chen, Cornelius, Martin, and Chau}{Chen
  et~al\mbox{.}}{2018}]%
        {chen2018robust}
\bibfield{author}{\bibinfo{person}{Shang-Tse Chen}, \bibinfo{person}{Cory
  Cornelius}, \bibinfo{person}{Jason Martin}, {and} \bibinfo{person}{Duen~Horng
  Chau}.} \bibinfo{year}{2018}\natexlab{}.
\newblock \showarticletitle{Robust Physical Adversarial Attack on Faster
  {R-CNN} Object Detector}.
\newblock \bibinfo{journal}{\emph{ArXiv:1804.05810}} (\bibinfo{year}{2018}).
\newblock


\bibitem[\protect\citeauthoryear{Chou, Tram{\`e}r, Pellegrino, and Boneh}{Chou
  et~al\mbox{.}}{2018}]%
        {chou2018sentinet}
\bibfield{author}{\bibinfo{person}{Edward Chou}, \bibinfo{person}{Florian
  Tram{\`e}r}, \bibinfo{person}{Giancarlo Pellegrino}, {and}
  \bibinfo{person}{Dan Boneh}.} \bibinfo{year}{2018}\natexlab{}.
\newblock \showarticletitle{{Sentinet}: Detecting physical attacks against deep
  learning systems}.
\newblock \bibinfo{journal}{\emph{arXiv:1812.00292}} (\bibinfo{year}{2018}).
\newblock


\bibitem[\protect\citeauthoryear{Feinman, Curtin, Shintre, and Gardner}{Feinman
  et~al\mbox{.}}{2017}]%
        {feinman2017detecting}
\bibfield{author}{\bibinfo{person}{Reuben Feinman}, \bibinfo{person}{Ryan~R.
  Curtin}, \bibinfo{person}{Saurabh Shintre}, {and} \bibinfo{person}{Andrew~B.
  Gardner}.} \bibinfo{year}{2017}\natexlab{}.
\newblock \bibinfo{title}{Detecting Adversarial Samples from Artifacts}.
\newblock
\newblock


\bibitem[\protect\citeauthoryear{Geiger, Lenz, and Urtasun}{Geiger
  et~al\mbox{.}}{2012}]%
        {geiger2012we}
\bibfield{author}{\bibinfo{person}{Andreas Geiger}, \bibinfo{person}{Philip
  Lenz}, {and} \bibinfo{person}{Raquel Urtasun}.}
  \bibinfo{year}{2012}\natexlab{}.
\newblock \showarticletitle{Are we ready for autonomous driving? the kitti
  vision benchmark suite}. In \bibinfo{booktitle}{\emph{Proceedings of the IEEE
  Conference on Computer Vision and Pattern Recognition (CVPR)}}.
\newblock


\bibitem[\protect\citeauthoryear{Greenberg}{Greenberg}{2018}]%
        {greenberg2018jeep}
\bibfield{author}{\bibinfo{person}{Andy Greenberg}.}
  \bibinfo{year}{2018}\natexlab{}.
\newblock \bibinfo{title}{Hackers Remotely Kill a Jeep on the Highway-With Me
  in It}.
\newblock
\newblock
\urldef\tempurl%
\url{https://www.wired.com/2015/07/hackers-remotely-kill-jeep-highway/}
\showURL{%
\tempurl}


\bibitem[\protect\citeauthoryear{Grosse, Manoharan, Papernot, Backes, and
  McDaniel}{Grosse et~al\mbox{.}}{2017}]%
        {grosse2017statistical}
\bibfield{author}{\bibinfo{person}{Kathrin Grosse}, \bibinfo{person}{Praveen
  Manoharan}, \bibinfo{person}{Nicolas Papernot}, \bibinfo{person}{Michael
  Backes}, {and} \bibinfo{person}{Patrick McDaniel}.}
  \bibinfo{year}{2017}\natexlab{}.
\newblock \showarticletitle{On the (statistical) detection of adversarial
  examples}.
\newblock \bibinfo{journal}{\emph{arXiv:1702.06280}}.
\newblock


\bibitem[\protect\citeauthoryear{Hinton, Deng, Yu, Dahl, Mohamed, Jaitly,
  Senior, Vanhoucke, Nguyen, Sainath, et~al\mbox{.}}{Hinton
  et~al\mbox{.}}{2012}]%
        {hinton2012deep}
\bibfield{author}{\bibinfo{person}{Geoffrey Hinton}, \bibinfo{person}{Li Deng},
  \bibinfo{person}{Dong Yu}, \bibinfo{person}{George~E Dahl},
  \bibinfo{person}{Abdel-rahman Mohamed}, \bibinfo{person}{Navdeep Jaitly},
  \bibinfo{person}{Andrew Senior}, \bibinfo{person}{Vincent Vanhoucke},
  \bibinfo{person}{Patrick Nguyen}, \bibinfo{person}{Tara~N Sainath},
  {et~al\mbox{.}}} \bibinfo{year}{2012}\natexlab{}.
\newblock \showarticletitle{Deep neural networks for acoustic modeling in
  speech recognition: The shared views of four research groups}. In
  \bibinfo{booktitle}{\emph{IEEE Signal Processing Magazine}}.
\newblock


\bibitem[\protect\citeauthoryear{Hinton, Sabour, and Frosst}{Hinton
  et~al\mbox{.}}{2018}]%
        {hinton2018matrix}
\bibfield{author}{\bibinfo{person}{Geoffrey Hinton}, \bibinfo{person}{Sara
  Sabour}, {and} \bibinfo{person}{Nicholas~and Frosst}.}
  \bibinfo{year}{2018}\natexlab{}.
\newblock \showarticletitle{Matrix capsules with EM routing}. In
  \bibinfo{booktitle}{\emph{International Conference on Learning
  Representations (ICLR)}}.
\newblock


\bibitem[\protect\citeauthoryear{Hinton, Krizhevsky, and Wang}{Hinton
  et~al\mbox{.}}{2011}]%
        {hinton2011transforming}
\bibfield{author}{\bibinfo{person}{Geoffrey~E Hinton}, \bibinfo{person}{Alex
  Krizhevsky}, {and} \bibinfo{person}{Sida~D Wang}.}
  \bibinfo{year}{2011}\natexlab{}.
\newblock \showarticletitle{Transforming auto-encoders}. In
  \bibinfo{booktitle}{\emph{International Conference on Artificial Neural
  Networks}}.
\newblock


\bibitem[\protect\citeauthoryear{Jia, Lu, Shen, Chen, Zhong, and Wei}{Jia
  et~al\mbox{.}}{2019}]%
        {jia2019fooling}
\bibfield{author}{\bibinfo{person}{Yunhan Jia}, \bibinfo{person}{Yantao Lu},
  \bibinfo{person}{Junjie Shen}, \bibinfo{person}{Qi~Alfred Chen},
  \bibinfo{person}{Zhenyu Zhong}, {and} \bibinfo{person}{Tao Wei}.}
  \bibinfo{year}{2019}\natexlab{}.
\newblock \showarticletitle{Fooling Detection Alone is Not Enough: First
  Adversarial Attack against Multiple Object Tracking}.
\newblock \bibinfo{journal}{\emph{arXiv:1905.11026}} (\bibinfo{year}{2019}).
\newblock


\bibitem[\protect\citeauthoryear{Kingma and Ba}{Kingma and Ba}{2014}]%
        {kingma2014adam}
\bibfield{author}{\bibinfo{person}{Diederik~P. Kingma} {and}
  \bibinfo{person}{Jimmy Ba}.} \bibinfo{year}{2014}\natexlab{}.
\newblock \bibinfo{title}{{Adam}: A Method for Stochastic Optimization}.
\newblock
\newblock


\bibitem[\protect\citeauthoryear{Kosiorek, Sabour, Teh, and Hinton}{Kosiorek
  et~al\mbox{.}}{2019}]%
        {kosiorek2019stacked}
\bibfield{author}{\bibinfo{person}{Adam Kosiorek}, \bibinfo{person}{Sara
  Sabour}, \bibinfo{person}{Yee~Whye Teh}, {and} \bibinfo{person}{Geoffrey~E
  Hinton}.} \bibinfo{year}{2019}\natexlab{}.
\newblock \showarticletitle{Stacked capsule autoencoders}. In
  \bibinfo{booktitle}{\emph{Advances in Neural Information Processing Systems
  (NeurIPS)}}.
\newblock


\bibitem[\protect\citeauthoryear{Krizhevsky, Sutskever, and Hinton}{Krizhevsky
  et~al\mbox{.}}{2012}]%
        {krizhevsky2012imagenet}
\bibfield{author}{\bibinfo{person}{Alex Krizhevsky}, \bibinfo{person}{Ilya
  Sutskever}, {and} \bibinfo{person}{Geoffrey~E Hinton}.}
  \bibinfo{year}{2012}\natexlab{}.
\newblock \showarticletitle{Imagenet classification with deep convolutional
  neural networks}. In \bibinfo{booktitle}{\emph{Advances in Neural Information
  Processing Systems (NeurIPS)}}.
\newblock


\bibitem[\protect\citeauthoryear{Kurakin, Goodfellow, and Bengio}{Kurakin
  et~al\mbox{.}}{2016}]%
        {kurakin2016adversarial}
\bibfield{author}{\bibinfo{person}{Alexey Kurakin}, \bibinfo{person}{Ian
  Goodfellow}, {and} \bibinfo{person}{Samy Bengio}.}
  \bibinfo{year}{2016}\natexlab{}.
\newblock \showarticletitle{Adversarial examples in the physical world}.
\newblock \bibinfo{journal}{\emph{arXiv:1607.02533}} (\bibinfo{year}{2016}).
\newblock


\bibitem[\protect\citeauthoryear{LaLonde and Bagci}{LaLonde and Bagci}{2018}]%
        {lalonde2018capsules}
\bibfield{author}{\bibinfo{person}{Rodney LaLonde} {and} \bibinfo{person}{Ulas
  Bagci}.} \bibinfo{year}{2018}\natexlab{}.
\newblock \showarticletitle{Capsules for object segmentation}.
\newblock \bibinfo{journal}{\emph{arXiv:1804.04241}} (\bibinfo{year}{2018}).
\newblock


\bibitem[\protect\citeauthoryear{Lee and Kolter}{Lee and Kolter}{2019}]%
        {lee2019physical}
\bibfield{author}{\bibinfo{person}{Mark Lee} {and} \bibinfo{person}{Zico
  Kolter}.} \bibinfo{year}{2019}\natexlab{}.
\newblock \showarticletitle{On physical adversarial patches for object
  detection}.
\newblock \bibinfo{journal}{\emph{arXiv:1906.11897}} (\bibinfo{year}{2019}).
\newblock


\bibitem[\protect\citeauthoryear{Lillicrap, Hunt, Pritzel, Heess, Erez, Tassa,
  Silver, and Wierstra}{Lillicrap et~al\mbox{.}}{2015}]%
        {lillicrap2015continuous}
\bibfield{author}{\bibinfo{person}{Timothy~P Lillicrap},
  \bibinfo{person}{Jonathan~J Hunt}, \bibinfo{person}{Alexander Pritzel},
  \bibinfo{person}{Nicolas Heess}, \bibinfo{person}{Tom Erez},
  \bibinfo{person}{Yuval Tassa}, \bibinfo{person}{David Silver}, {and}
  \bibinfo{person}{Daan Wierstra}.} \bibinfo{year}{2015}\natexlab{}.
\newblock \showarticletitle{Continuous control with deep reinforcement
  learning}.
\newblock \bibinfo{journal}{\emph{arXiv:1509.02971}} (\bibinfo{year}{2015}).
\newblock


\bibitem[\protect\citeauthoryear{Liu, Ma, Aafer, Lee, Zhai, Wang, and
  Zhang}{Liu et~al\mbox{.}}{2017}]%
        {liu2017trojaning}
\bibfield{author}{\bibinfo{person}{Yingqi Liu}, \bibinfo{person}{Shiqing Ma},
  \bibinfo{person}{Yousra Aafer}, \bibinfo{person}{Wen-Chuan Lee},
  \bibinfo{person}{Juan Zhai}, \bibinfo{person}{Weihang Wang}, {and}
  \bibinfo{person}{Xiangyu Zhang}.} \bibinfo{year}{2017}\natexlab{}.
\newblock \showarticletitle{Trojaning attack on neural networks}.
\newblock  (\bibinfo{year}{2017}).
\newblock


\bibitem[\protect\citeauthoryear{Lu, Issaranon, and Forsyth}{Lu
  et~al\mbox{.}}{2017}]%
        {lu2017quantized}
\bibfield{author}{\bibinfo{person}{Jiajun Lu}, \bibinfo{person}{Theerasit
  Issaranon}, {and} \bibinfo{person}{David Forsyth}.}
  \bibinfo{year}{2017}\natexlab{}.
\newblock \showarticletitle{SafetyNet: Detecting and Rejecting Adversarial
  Examples Robustly}. In \bibinfo{booktitle}{\emph{The IEEE International
  Conference on Computer Vision (ICCV)}}.
\newblock


\bibitem[\protect\citeauthoryear{Meng and Chen}{Meng and Chen}{2017}]%
        {meng2017magnet}
\bibfield{author}{\bibinfo{person}{Dongyu Meng} {and} \bibinfo{person}{Hao
  Chen}.} \bibinfo{year}{2017}\natexlab{}.
\newblock \showarticletitle{{MagNet}: a two-pronged defense against adversarial
  examples}. In \bibinfo{booktitle}{\emph{ACM SIGSAC Conference on Computer and
  Communications Security (CCS)}}.
\newblock


\bibitem[\protect\citeauthoryear{Metzen, Genewein, Fischer, and
  Bischoff}{Metzen et~al\mbox{.}}{2017}]%
        {metzen2017detecting}
\bibfield{author}{\bibinfo{person}{Jan~Hendrik Metzen}, \bibinfo{person}{Tim
  Genewein}, \bibinfo{person}{Volker Fischer}, {and} \bibinfo{person}{Bastian
  Bischoff}.} \bibinfo{year}{2017}\natexlab{}.
\newblock \showarticletitle{On detecting adversarial perturbations}.
\newblock \bibinfo{journal}{\emph{International Conference on Learning
  Representations (ICLR)}}.
\newblock


\bibitem[\protect\citeauthoryear{Moosavi-Dezfooli, Fawzi, and
  Frossard}{Moosavi-Dezfooli et~al\mbox{.}}{2016}]%
        {moosavi2016deepfool}
\bibfield{author}{\bibinfo{person}{Seyed-Mohsen Moosavi-Dezfooli},
  \bibinfo{person}{Alhussein Fawzi}, {and} \bibinfo{person}{Pascal Frossard}.}
  \bibinfo{year}{2016}\natexlab{}.
\newblock \showarticletitle{{Deepfool}: a simple and accurate method to fool
  deep neural networks}. In \bibinfo{booktitle}{\emph{Proceedings of the IEEE
  Conference on Computer Vision and Pattern Recognition (CVPR)}}.
\newblock


\bibitem[\protect\citeauthoryear{Pang, Du, Dong, and Zhu}{Pang
  et~al\mbox{.}}{2017}]%
        {pang2017robust}
\bibfield{author}{\bibinfo{person}{Tianyu Pang}, \bibinfo{person}{Chao Du},
  \bibinfo{person}{Yinpeng Dong}, {and} \bibinfo{person}{Jun Zhu}.}
  \bibinfo{year}{2017}\natexlab{}.
\newblock \showarticletitle{Towards Robust Detection of Adversarial Examples}.
\newblock \bibinfo{journal}{\emph{arXiv:1706.00633}} (\bibinfo{year}{2017}).
\newblock


\bibitem[\protect\citeauthoryear{Papernot, McDaniel, and Goodfellow}{Papernot
  et~al\mbox{.}}{2016}]%
        {papernot2016transferability}
\bibfield{author}{\bibinfo{person}{Nicolas Papernot}, \bibinfo{person}{Patrick
  McDaniel}, {and} \bibinfo{person}{Ian Goodfellow}.}
  \bibinfo{year}{2016}\natexlab{}.
\newblock \showarticletitle{Transferability in machine learning: from phenomena
  to black-box attacks using adversarial samples}.
\newblock \bibinfo{journal}{\emph{arXiv:1605.07277}} (\bibinfo{year}{2016}).
\newblock


\bibitem[\protect\citeauthoryear{Qin, Frosst, Sabour, Raffel, Cottrell, and
  Hinton}{Qin et~al\mbox{.}}{2019}]%
        {qin2019detecting}
\bibfield{author}{\bibinfo{person}{Yao Qin}, \bibinfo{person}{Nicholas Frosst},
  \bibinfo{person}{Sara Sabour}, \bibinfo{person}{Colin Raffel},
  \bibinfo{person}{Garrison Cottrell}, {and} \bibinfo{person}{Geoffrey
  Hinton}.} \bibinfo{year}{2019}\natexlab{}.
\newblock \showarticletitle{Detecting and diagnosing adversarial images with
  class-conditional capsule reconstructions}.
\newblock \bibinfo{journal}{\emph{arXiv:1907.02957}} (\bibinfo{year}{2019}).
\newblock


\bibitem[\protect\citeauthoryear{Sabour, Frosst, and Hinton}{Sabour
  et~al\mbox{.}}{2017}]%
        {sabour2017dynamic}
\bibfield{author}{\bibinfo{person}{Sara Sabour}, \bibinfo{person}{Nicholas
  Frosst}, {and} \bibinfo{person}{Geoffrey~E Hinton}.}
  \bibinfo{year}{2017}\natexlab{}.
\newblock \showarticletitle{Dynamic routing between capsules}. In
  \bibinfo{booktitle}{\emph{Advances in Neural Information Processing Systems
  (NeurIPS)}}.
\newblock


\bibitem[\protect\citeauthoryear{Srivastava, Hinton, Krizhevsky, Sutskever, and
  Salakhutdinov}{Srivastava et~al\mbox{.}}{2014}]%
        {srivastava2014dropout}
\bibfield{author}{\bibinfo{person}{Nitish Srivastava},
  \bibinfo{person}{Geoffrey Hinton}, \bibinfo{person}{Alex Krizhevsky},
  \bibinfo{person}{Ilya Sutskever}, {and} \bibinfo{person}{Ruslan
  Salakhutdinov}.} \bibinfo{year}{2014}\natexlab{}.
\newblock \showarticletitle{{Dropout}: a simple way to prevent neural networks
  from overfitting}.
\newblock \bibinfo{journal}{\emph{The Journal of Machine Learning Research}}
  (\bibinfo{year}{2014}).
\newblock


\bibitem[\protect\citeauthoryear{Szegedy, Zaremba, Sutskever, Bruna, Erhan,
  Goodfellow, and Fergus}{Szegedy et~al\mbox{.}}{2013}]%
        {szegedy2013intriguing}
\bibfield{author}{\bibinfo{person}{Christian Szegedy},
  \bibinfo{person}{Wojciech Zaremba}, \bibinfo{person}{Ilya Sutskever},
  \bibinfo{person}{Joan Bruna}, \bibinfo{person}{Dumitru Erhan},
  \bibinfo{person}{Ian Goodfellow}, {and} \bibinfo{person}{Rob Fergus}.}
  \bibinfo{year}{2013}\natexlab{}.
\newblock \showarticletitle{Intriguing properties of neural networks}.
\newblock \bibinfo{journal}{\emph{arXiv:1312.6199}} (\bibinfo{year}{2013}).
\newblock


\bibitem[\protect\citeauthoryear{Thys, Van~Ranst, and Goedem{\'e}}{Thys
  et~al\mbox{.}}{2019}]%
        {thys2019fooling}
\bibfield{author}{\bibinfo{person}{Simen Thys}, \bibinfo{person}{Wiebe
  Van~Ranst}, {and} \bibinfo{person}{Toon Goedem{\'e}}.}
  \bibinfo{year}{2019}\natexlab{}.
\newblock \showarticletitle{Fooling automated surveillance cameras: adversarial
  patches to attack person detection}. In \bibinfo{booktitle}{\emph{Proceedings
  of the IEEE Conference on Computer Vision and Pattern Recognition (CVPR)
  Workshops}}.
\newblock


\bibitem[\protect\citeauthoryear{Wan, Zeiler, Zhang, Le~Cun, and Fergus}{Wan
  et~al\mbox{.}}{2013}]%
        {wan2013regularization}
\bibfield{author}{\bibinfo{person}{Li Wan}, \bibinfo{person}{Matthew Zeiler},
  \bibinfo{person}{Sixin Zhang}, \bibinfo{person}{Yann Le~Cun}, {and}
  \bibinfo{person}{Rob Fergus}.} \bibinfo{year}{2013}\natexlab{}.
\newblock \showarticletitle{Regularization of neural networks using
  dropconnect}. In \bibinfo{booktitle}{\emph{International Conference on
  Machine Learning (ICML)}}.
\newblock


\bibitem[\protect\citeauthoryear{Weisstein}{Weisstein}{2004}]%
        {weisstein2004affine}
\bibfield{author}{\bibinfo{person}{Eric~W Weisstein}.}
  \bibinfo{year}{2004}\natexlab{}.
\newblock \showarticletitle{Affine transformation}.
\newblock \bibinfo{journal}{\emph{https://mathworld. wolfram. com/}}
  (\bibinfo{year}{2004}).
\newblock


\bibitem[\protect\citeauthoryear{Xu, Zhang, Liu, Fan, Sun, Chen, Chen, Wang,
  and Lin}{Xu et~al\mbox{.}}{2019}]%
        {xu2019adversarial}
\bibfield{author}{\bibinfo{person}{Kaidi Xu}, \bibinfo{person}{Gaoyuan Zhang},
  \bibinfo{person}{Sijia Liu}, \bibinfo{person}{Quanfu Fan},
  \bibinfo{person}{Mengshu Sun}, \bibinfo{person}{Hongge Chen},
  \bibinfo{person}{Pin-Yu Chen}, \bibinfo{person}{Yanzhi Wang}, {and}
  \bibinfo{person}{Xue Lin}.} \bibinfo{year}{2019}\natexlab{}.
\newblock \showarticletitle{Adversarial T-shirt! Evading Person Detectors in A
  Physical World}.
\newblock \bibinfo{journal}{\emph{arXiv:1910.11099}} (\bibinfo{year}{2019}).
\newblock


\bibitem[\protect\citeauthoryear{Xu, Evans, and Qi}{Xu et~al\mbox{.}}{2018}]%
        {xu2018feature}
\bibfield{author}{\bibinfo{person}{Weilin Xu}, \bibinfo{person}{David Evans},
  {and} \bibinfo{person}{Yanjun Qi}.} \bibinfo{year}{2018}\natexlab{}.
\newblock \showarticletitle{Feature Squeezing: Detecting Adversarial Examples
  in Deep Neural Networks}.
\newblock \bibinfo{journal}{\emph{Proceedings 2018 Network and Distributed
  System Security Symposium}} (\bibinfo{year}{2018}).
\newblock
\showISBNx{1891562495}
\urldef\tempurl%
\url{https://doi.org/10.14722/ndss.2018.23198}
\showDOI{\tempurl}


\bibitem[\protect\citeauthoryear{Yu, Hu, Guo, Chao, and Weinberger}{Yu
  et~al\mbox{.}}{2019}]%
        {yu2019weakness}
\bibfield{author}{\bibinfo{person}{Tao Yu}, \bibinfo{person}{Shengyuan Hu},
  \bibinfo{person}{Chuan Guo}, \bibinfo{person}{Wei-Lun Chao}, {and}
  \bibinfo{person}{Kilian~Q. Weinberger}.} \bibinfo{year}{2019}\natexlab{}.
\newblock \showarticletitle{A New Defense Against Adversarial Images: Turning a
  Weakness into a Strength}.
\newblock \bibinfo{journal}{\emph{arXiv:1910.07629}} (\bibinfo{year}{2019}).
\newblock


\end{thebibliography}
